# What can we learn from marine geophysics to study rifted margins?

Julia Autin[1] and Louise Watremez[2]

1. Université de Strasbourg, CNRS, Institut Terre et Environnement de Strasbourg, UMR 7063, 5 rue Descartes, Strasbourg F-67084, France
2. Univ. Lille, CNRS, Univ. Littoral Côte d'Opale, UMR 8187, LOG, Laboratoire d'Océanologie et de Géosciences, F-59000 Lille, France

# A. Geophysical methods

Present-day continental rifted margins are mostly located under water. Access to rocks and structures in the offshore is possible with drillings or dredges, but the acquisition of such data types is very expensive and time consuming, only leading to local constraints. Consequently, such scarce and punctual datasets do not allow obtaining a global picture of the structure of rifted margins. To acquire such knowledge, researchers may rather use indirect data, such as geophysical methods.

Below, we list and briefly describe the main geophysical methods used when studying offshore rifts and rifted margins. If interested, and for further details and explanations, the reader is referred to contributions specifically dedicated to the methods. A list of references is proposed at the end of this chapter.

## *1. Seismic imaging*

Several methods are based on the properties of seismic waves, i.e. acoustic waves, as they propagate through the Earth layers. However, the observation scale varies with the method. For example, with increasing depth of observation, users can choose between sub-bottom profilers, sparker, boomer, single-channel seismic, multichannel deep seismic, or seismic refraction data. Here, we focus on the deep-penetrating methods, allowing for crustal scale observations: multichannel deep seismic and seismic refraction.

The seismic energy propagates from the source, down to the seafloor and geological layers. It is partly reflected or refracted at each geological boundary (interface), depending on the layers' impedance contrast (Figure 1). This impedance contrast corresponds to a change in physical properties (seismic velocities and densities). However, it is often difficult to determine why such changes occur: is it a lithological change (a different kind of rock media), a discontinuity in rock media (e.g., faults, intrusions), a change in fluid content in the material, etc.? Moreover, the imaging can generate artefacts, i.e., element in the underground image that are not real but inherent to the behavior of the acoustic waves (lateral echoes, multiple arrivals, slope effects, diffraction hyperbolas, background noise…). Thus, it is of primary importance to separate observation of the seismic data from their interpretation. Observation should reach a consensus amongst the researchers whereas interpretation is more subjective, and ideally based on discussion and comparison with other datasets.

### *a) Multichannel deep seismic*

- *Acquisition*

The acquisition of offshore deep seismic is done either in 2D (along the path of the ship) or in 3D (with multiple receiver arrays along and laterally to the ship path). The geometry of the seismic source and receivers is set to maintain a constant distance between the source and each receiver (offset, Figure 1a). This favors the observation of reflected wave arrivals instead of refracted wave arrivals, which usually occur at larger offsets. The receivers are aligned along one (2D) or several streamers (3D), which lengths may be up to 10-12 km, for the best configurations. The source is chosen depending on the target. Most of the time, scientists make a compromise between penetration (large source volume) and resolution (smaller sources) as they need both energy to penetrate down to the crustal basement and sufficient resolution in the sediments (~10 m).

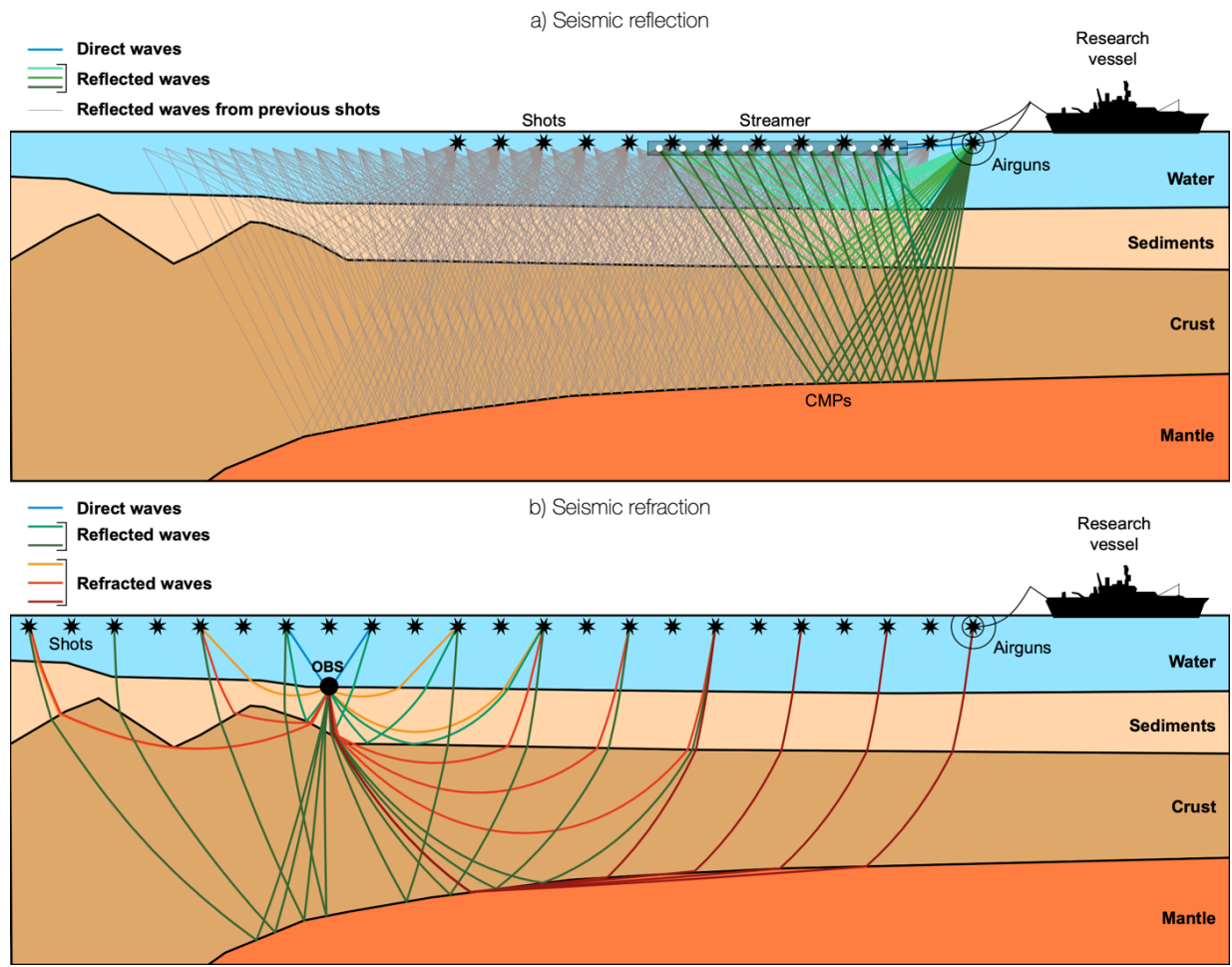

*Figure 1: Acquisition principles of seismic reflection and refraction. The research vessel's outline belongs to a French ship, the N/O L'Atalante. a) For seismic reflection, the constant source-receiver distance allows the record of near-vertical reflected waves and the multiple receivers and shots sample common mid-points (CMPs). See the text for further explanations. (b) Seismic refraction is based on increasing source-receiver distances, creating a wider incidence angle and allowing to record deeper refracted waves as the offsets increase.*

The quality of the final seismic profiles highly depends on the fold, i.e., the theoretical number of times a given point of the seafloor (CMP: common mid-point) is sampled by a reflection and recorded by a receiver (Figure 2). The greater the fold, the better the signal to noise ratio. The fold depends on the number of receivers, their spacing, and the shot interval distance (e.g. Yilmaz 2001).

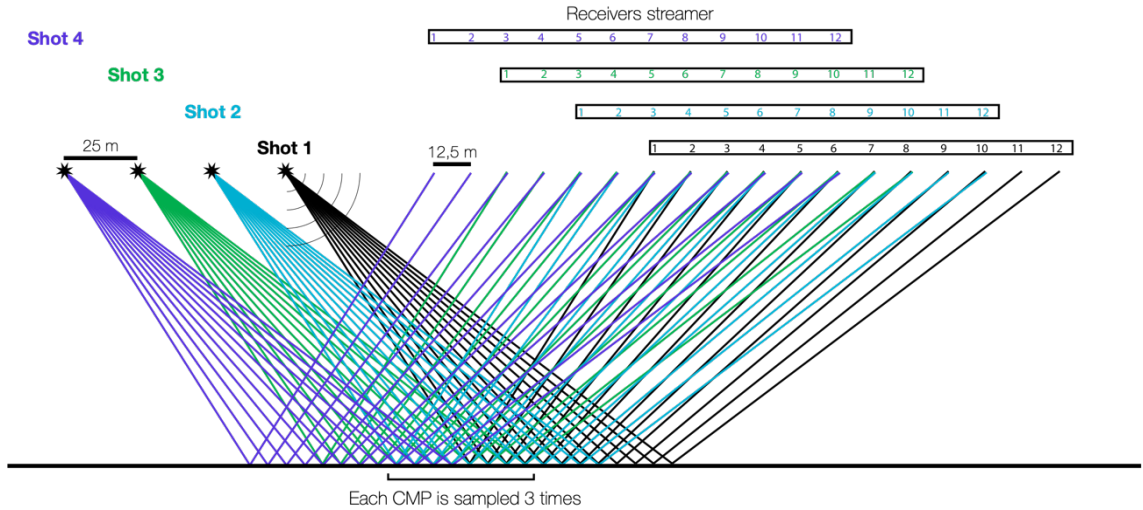

*Figure 2: Common mid-points (CMP) acquisition. Each shot is recorded by the 12 receivers of the streamer. A same point of the seafloor (CMP) is sampled several times by several shot/receiver pairs. Here each CMP is sampled 3 times. The 3 records are stacked together to improve the signal to noise ratio.*

- Data processing

Seismic processing aims at reducing the signal to noise ratio. Noise can be related to known behavior of the seismic waves (multiples, diffraction hyperbola, etc., see Figure 3) or to disturbances during the acquisition (wind, currents, human activity, internal Earth activity, source or receiver issues…).

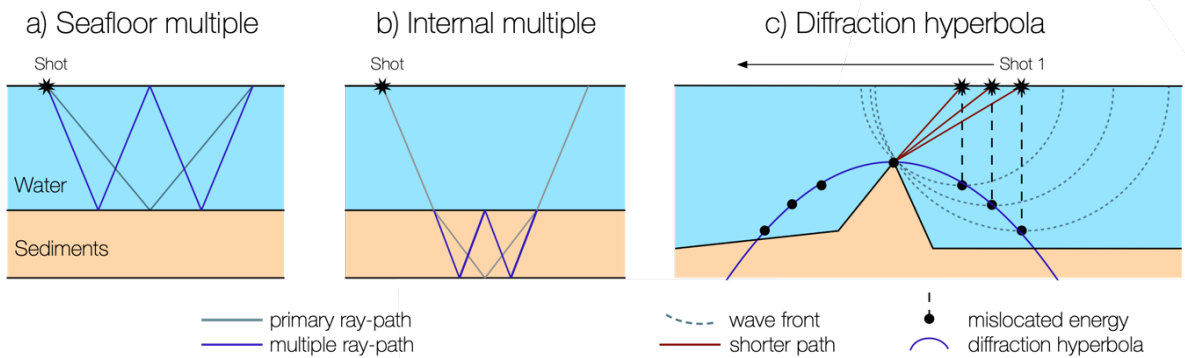

*Figure 3: Noises related to seismic waves behavior. (a) The seafloor multiple results from waves reflected first on the seafloor then under the sea surface and finally on the seafloor again. It produces a ghost of the seafloor with arrival-times doubled compared to the seafloor arrivals. (b) Similar behavior within a geological layer can produce internal multiples as well. (c) Diffraction hyperbola result from diffraction points, such as faults: the seismic energy is mis-located because the diffraction point is reached faster laterally than the actual interface, vertically (the mis-located energy forms a hyperbola).*

The quality control of the traces allows to select or improve the seismic records (filtering and offset- or time-dependent gain applied on the data, either as shot gathers or as receiver gathers). The source deconvolution aims at reducing the length of the source wavelet to its minimum in order to obtain a sharper signal and improve the resolution. Traces are sorted as common midpoint gathers (CMP), i.e. all the traces recording a theoretical point of the seafloor (Figure 2). CMP gathers display the arrival times at a common spatial point as a function of the offset (source-receiver distance) and shows a downward hyperbolic shape (Figure 4a). Normal move out (NMO) consists in the correction of the arrival times with a theoretical velocity (stacking velocity) in order to align each arrival at the same level (Figure 4b). Such process assumes that the reflector corresponding to those arrivals is horizontal. Stacking consists in summing the energy recorded by each NMO-corrected trace to reinforce the signal in a given point of the underground, and thus improving the signal to noise ratio (Figure 4c). It is particularly efficient to reduce multiples as their arrival-times are corrected with a higher stacking velocity than needed, resulting in downward alignment after NMO and thus avoiding the stacking of the multiple energy.

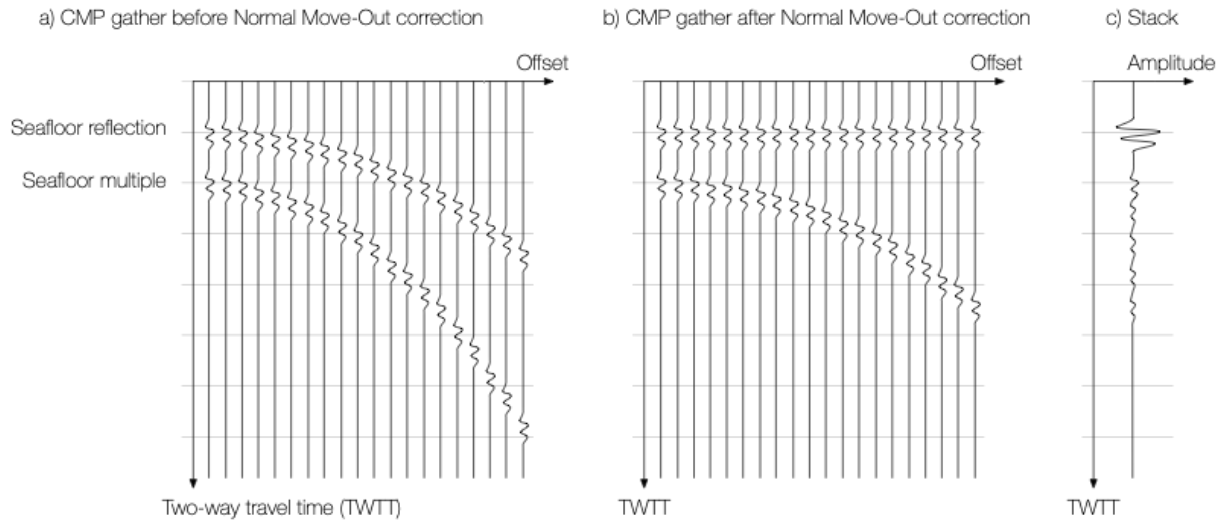

*Figure 4: Principle of normal move-out (NMO) correction before stacking a) CMP gather before NMO, b) CMP gather after NMO, c) stacked CMPs.*

After stacking, the seismic section is obtained by juxtaposing the CMP stacked traces along the profile. Further processing is operated to improve the final seismic section. For instance, the migration of the data allows to remove diffraction hyperbolas (Figure 3c). The aim of the migration is to restitute the energy at the apex of the hyperbola, in order to place the signal at the right geometrical point (Figure 5). Slope migration corrects the effect of non-horizontal reflections: in this case the theoretical CMP is not focused on one point on the reflector but scattered around it.

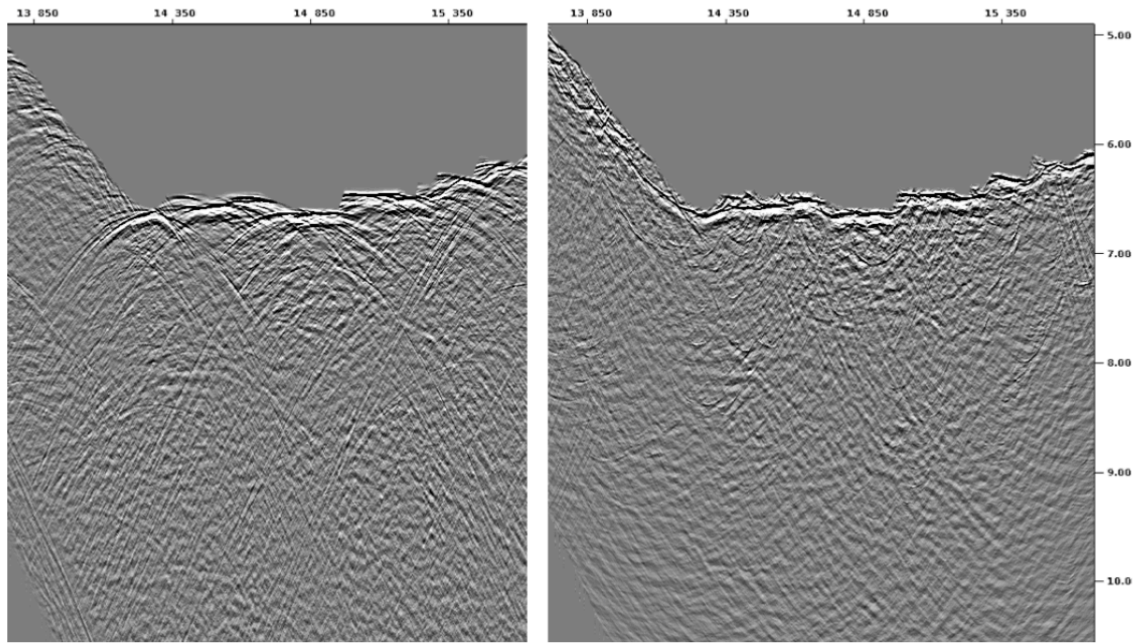

*Figure 5: Effect of migration at constant velocity (water velocity,1500 m/sec) on a stacked section (cruise report, Leroy, Cannat 2014). Left: stacked section. Right: migrated section.*

Finally, time-depth conversion is useful to assess the real geometry of the cross-section. It needs a thorough estimation of the propagation velocities and as such should be considered as a model rather than a result.

We presented the basic steps of processing. Of course, advanced processing includes more thorough analyses depending on the targets and specificities of the acquisition. For example, more advanced antimultiples will be applied if the target is deeper than the multiple time arrival, Deep Move Out (DMO) is applied where steep slopes are present, Pre-Stack Depth Migration (PSDM) is a more elaborate way to convert data from time to depth. We encourage the reader to study dedicated literature for further details (e.g. Yilmaz 2001; Dondurur 2018).

- Observation and interpretation

To allow an objective interpretation of seismic data we follow 3 main steps during their study: (1) the knowledge of the data acquisition and processing informs about the quality of the signal to noise ratio improvement, the elimination of artefacts and the accuracy of the eventual depth conversion. (2) Seismic observation allows to describe the image without preconception on the geological background. (3) Seismic interpretation allows to propose a possible architecture of the rifted margin through a scientific argumentation, which will remain theoretical in the absence of in situ data.

It is worth saying that beginners should acquire a good understanding of the processing steps required for the production of seismic profiles. Indeed, the observer needs to know whether the seismic section is in time or depth before considering geometrical aspects (angles, thicknesses, volumes…), if multiples still appear, if a migration was applied, etc. This allows to estimate the degree of confidence in the observations.

Then, seismic observation consists in defining seismic facies (i.e. packages of reflections that display a common seismic pattern) relative to seismic units (Figure 6). It can be defined using criteria such as amplitude, frequency, continuity, but also with geometrical observations describing the termination of seismic reflections (onlap, dowlap, toplap, truncation…) or their internal geometry (e.g. parallel, oceanward or continentward dipping, wedge-shaped, divergent, concave, folded reflections, affected by major or minor faults). Nevertheless, the internal geometry of a given seismic unit can vary along a seismic line. The goal is to achieve a coherent description of the seismic facies throughout the entire profile and then the whole seismic dataset.

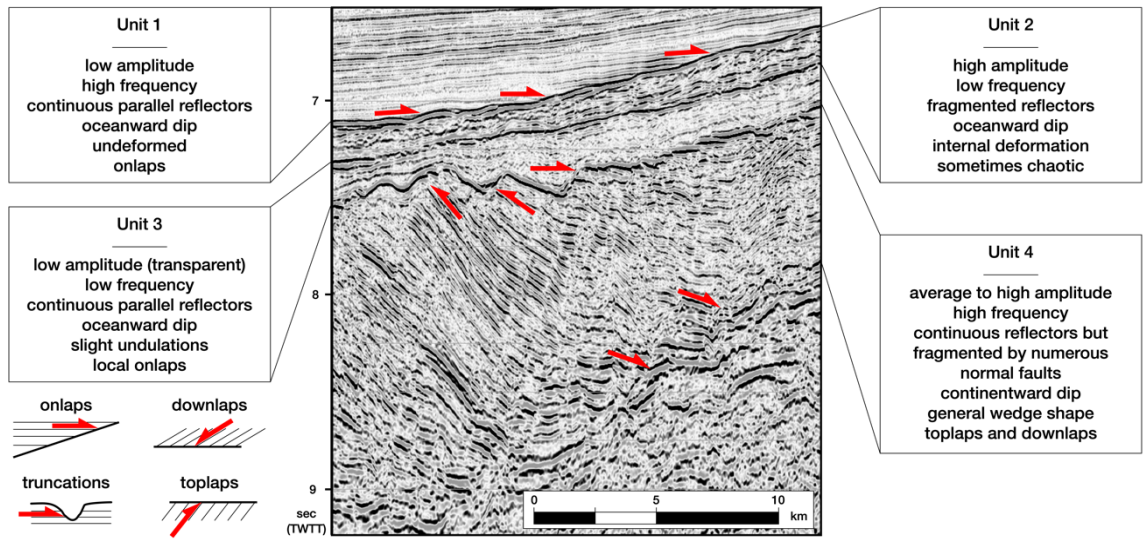

*Figure 6: Example of seismic facies description. Red arrows highlight the termination of reflectors in each unit.*

Finally, the actual interpretation begins. One should be able to recognize a logical pattern in deposition of the sedimentary seismic units from base to top, in order to constrain a depositional and tectonic history. If any well or dredge data are available, they bring further information on the nature and age of the rocks. This allows to point out possible gaps in the sedimentary column. For example, an observed seismic unit in the proximal domain can be missing in the distal domain, due to erosion (erosional surfaces could be observed), sediment bypass, non-deposition (e.g. due to delayed subsidence), or changes in the deposited material at a given time. Thick sand deposits in the proximal domain can evolve into thin condensed intervals in the distal domain. Hence, the evolution of the sedimentation (thicknesses, bypass, erosion…) gives indication on the available accommodation space for sedimentation and distance to sediment sources during rifting, and thus on the subsidence history.

One must study carefully the sedimentary units but also the basement seismic units. Indeed, it is sometime possible to recognize distinct basement seismic facies. For example, an oceanic basement can be distinguished from a continental basement by the strong top reflectivity, clear Moho reflection (crust typically about 2 s TWTT thick) and transparent facies of the magmatic oceanic crust (see also part B.1.a). Nevertheless, it will be difficult to distinguish between (1) a thin continental crust that is for instance deformed, highly intruded and covered by magmatic additions, (2) a zone of exhumed mantle that may also be highly intruded and covered by magmatic additions, or (3) a slightly tectonised magmatic oceanic crust (see also Chapter XXX). In this case, the lack of the basal sedimentary units could help to determine whether the basement is inherited (previous continental surface) or more recently exposed to the surface (exhumation of deeper crust or mantle, oceanic spreading). The presence or absence of Moho reflections is an indicator of the nature of the basement (see also part B.1.a). Indeed, exhumed mantle does not often display Moho reflections, due to progressive increase of the densities and velocities (no impedance contrast) with depth. Lower crust or oceanic crust should display a basal reflection (sharp impedance contrast).

### b) Seismic refraction

Seismic refraction allows to use the arrival times from different seismic phases (refracted and reflected, Figure 1b) in the geological layers, in order to build a velocity model of the subsurface, usually based on P-wave velocities. S-wave arrivals can also be used when P to S conversions occur and such arrivals are recorded (S-waves do not propagate in the water). Different modelling methods exist, showing different style of results (smooth or layered models) and using the datasets in different manners (first arrivals or more). It is important to understand how a seismic ray travels in the subsurface to be able to fathom the quality of a model, assess which areas are well-resolved (or not), and finally have an idea on how to interpret these (usually) colorful images.

- Principle

Seismic refraction, also called "refraction and wide-angle reflection" or just "wide-angle seismics" (R-WAR or WAS), uses the ray-tracing theory to assess seismic ray-paths in the subsurface, compute source-to-receiver travel-times and deduce the seismic velocities as well as depth and geometry of the geological interfaces. The simplistic theory, as usually taught, assumes simple layers of constant velocities, with head-waves propagating along the interfaces when the seismic velocities increase with depth. However, the head-wave corresponds to a very specific case, where the wave front hits the geological interface at an incidence angle such as the refracted angle is equal to 90° (and refracts along the interface). Most of the seismic energy is either refracted in the layer or reflected at the interface. Thus, head-waves show very low energy and the clear arrivals on the seismic refraction sections mostly belong to refracted waves (within layers) and wide-angle reflections. In reality, the seismic velocities in a geological layer tend to gradually increase with depth (vertical velocity gradient) and may also vary laterally (because of different compaction in the sediments, lateral variations in the nature of the geological structures, further lithostatic pressure increase in the crust or mineralogical changes with depth, etc.). One can imagine a geological layer as an infinity of very thin constant-velocity-layers where the seismic ray progressively refracts. Thus, the refracted waves usually turn within a geological layer until it comes back up to the receiver. Consequently, the understanding of the recorded travel-times involves modelling of these vertical and lateral variations through ray-tracing, using computer softwares (see data processing section).

- Acquisition

The acquisition of seismic refraction is achieved by recording seismic shots on multiple instruments, either along the profile (2D) or on a grid (3D). Seismic rays travel deeper in the Earth structures as the source-receiver distance increases (Figure 1b). Instruments are deployed from a research vessel at the beginning of the survey and are recovered once the seismic shots are done and recorded by the instruments. Seismic records are organized as receiver-gathers, with one trace per shot, and sorted by offsets (2D) or shot numbers (3D).

The data acquired during the scientific cruise are pre-processed in order to be ready for picking. The main pre-processing step consists in the instrument relocation. Indeed, Ocean-bottom seismometers (OBS) are deployed from the sea-surface, sink and drift, following currents down to the seafloor. As GPS signal does not transmit underwater, the exact position of the instrument has to be deduced by triangulation, with the help of the arrival-times of the direct waves, which propagated from the source to the instrument, in the water. Note that water depth and acoustic velocities must be known. An instrument that is properly relocated shows a symmetrical seismic record at short offsets (water-wave arrivals). The relocation process is essential to obtain accurate travel-times, and therefore, velocity modelling.

Some data filtering (e.g. low frequencies on hydrophone data) may be applied to the data, but it is best to pick on unfiltered data to avoid travel-time shifts due to the processing. Light bandpass filter, offset-dependent gain and deconvolution may be applied to help see far-offset arrivals (be careful, the filter may lead to a shift of the arrival-times) or data display.

- Data processing

Seismic refraction processing aims at building a velocity model of the crustal structures, down to the uppermost mantle, when the source-receiver distances are large enough. The velocity model should account for the observed travel-times on the seismic records. These arrival times are compared to computed arrival times, through ray-tracing in the model (Figure 7, middle and lower panels).

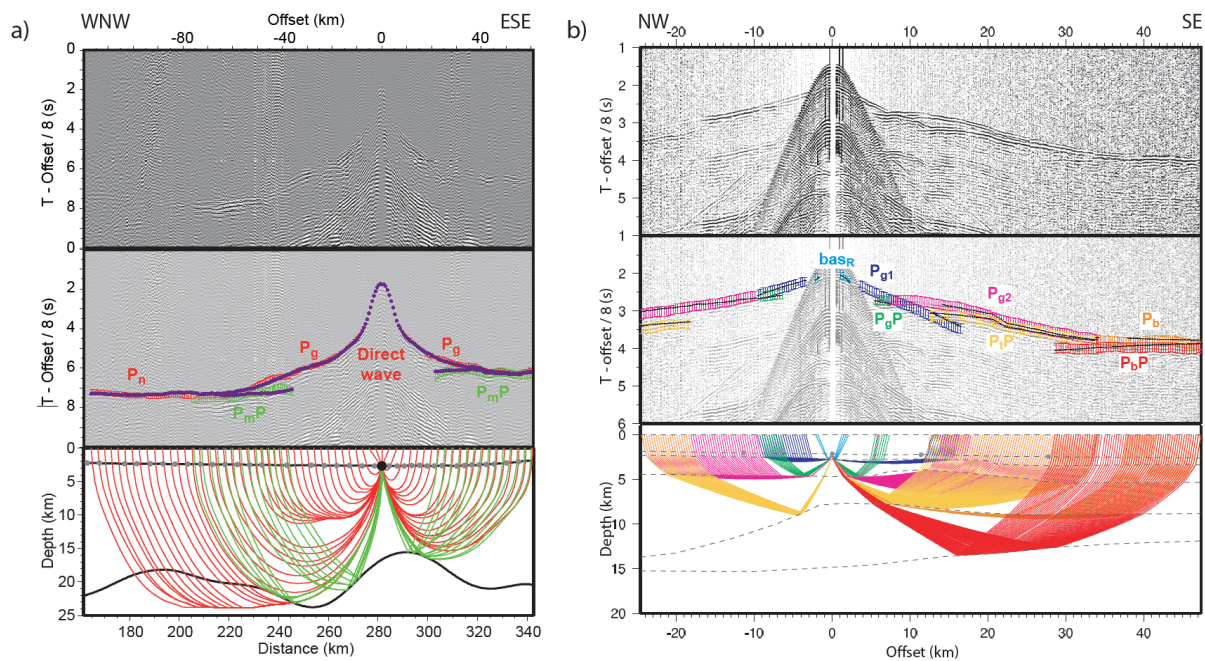

*Figure 7: Examples of data, picks and ray-tracing in tomography and forward layered models. a) Seismic record (top panel), picks, computed arrival times (middle panel) and rays (lower panel) for a tomography model (Orphan Basin, Watremez et al. 2015). All first arrival-times are picked as a unique phase and wide-angle reflections at the Moho are picked separately to model the depth of a reflector (black line). b) Same display for a layered forward model (North-Eastern Gulf of Aden, Watremez et al. 2011). Dashed lines show the model interfaces. Each seismic phase is picked separately, the code allows for choosing in which layer the ray travels, and whether it is to be traced as a refraction or a refraction. The color bars in the middle panels correspond to the picked arrival-times, their height being the uncertainty for each pick. The strong amplitude hyperbola at the center of the sections (centered around offset = 0 km) correspond to the direct waves that traveled in the water column (Figure 1b). Acronyms: $bas_R$ = top of acoustic basement reflection, $P_g$, $P_{g1}$, $P_{g2}$ = refracted arrivals in the crustal layers, $P_gP$ = mid-crustal reflection, $P_mP$ = Moho reflection, $P_tP$ and $P_bP$ = top and bottom of a lower-crustal body, respectively, $P_b$ = refraction in a lower-crustal body, and $P_n$ = upper-most mantle refractions. Purple (a) and black (b) dots correspond to the computed arrival times. The color codes of the picks and the rays in the lower panels are the same.*

Picking on the seismic records provides the seismic traveltimes used for the modelling. Depending on the modelling method, the data is picked following different strategies. Please note that the method and software examples mentioned below consist in some of the available softwares, used by the authors of this chapter, but do not consist in an exhaustive list of the existing softwares. The main seismic refraction modelling methods are (see Figure 8 for data picking and model examples along a same profile):

a) First arrival-time tomography (e.g. FAST, Zelt, Barton 1998): this inversion method only takes into account first arrival times and uses a velocity grid as input. The output is an updated smooth velocity grid, accounting for the travel-times. This method is interesting for onboard processing along 2D lines, for example, or 3D modelling.

b) Joint first-arrival time and wide-angle reflection tomography (e.g. Tomo2D, Korenaga *et al.* 2000): this other inversion method also uses first arrival-times (refracted arrivals), but a wide-angle picked dataset can also be added to the modelling, in order to obtain the depth and morphology of an interface (typically, the Moho discontinuity). This scheme may also be used for layered modelling, adding velocity jumps at interfaces, and modelling multiple interfaces, layer by layer (e.g. Sallarès *et al.* 2011). The result is a velocity grid (more or less smooth) with one or several interfaces (if modelled).

c) Forward layered modelling (e.g. Rayinvr, Zelt, Smith 1992): This method is used for 2D modelling, layer by layer, usually from top to bottom. It is more time consuming than the tomography methods but allows for slight geological interpretation during the modelling process. It allows for distinguishing in which layer each refracted ray turned, and which interface each reflected ray hit. The aim of this method is to explain the different seismic phases at best, first and secondary arrivals, using the least model parameters (number of layers and interface and velocity nodes) possible (Zelt 1999). It is possible to model the refracted arrival-times that occur at secondary arrivals, as well as S-wave arrivals, multiples, etc. The result consists in a layered velocity model, with sharp velocity discontinuities at model interfaces, generally related to lithological changes.

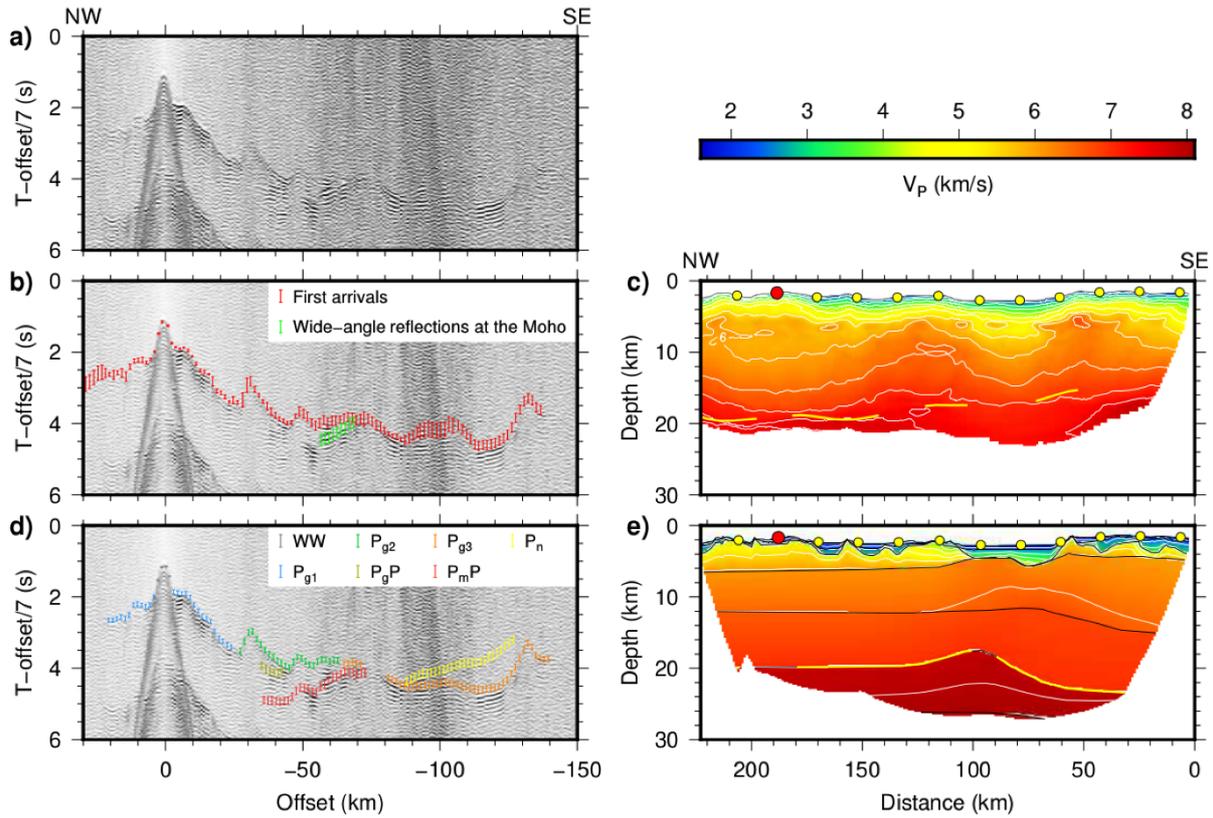

*Figure 8: Example of OBS data, picks and results along a same South China Sea profile using Tomo2D (Pichot et al. 2014) and Rayinvr (Liang et al. 2019). a) Example of seismic record along the profile. b) Same data with picks for the travel-time tomography. The color bars correspond to the picks, following the color code in the top-right of the panel. Every 10th picks are shown. Their heights correspond to the picking uncertainty. c) Velocity model resulting from the tomography inversion, using Tomo2D. The velocity color scale is shown on the top-right panel of the figure. The yellow lines correspond to the Moho depth obtained during the joint inversion, showing the interface where some reflected arrivals hit the interface. The white lines show iso-velocity contours, every 0.5 km/s. The yellow and red circles present the OBS locations along the bathymetry line, the red one corresponding to the data presented on the left. d) Same data as in a), with the picks used for forward layered modelling. Picks are presented the same way than in b). WW = direct wave arrivals ("water wave"), $P_{g1}$, $P_{g2}$ and $P_{g3}$ = refracted arrivals in the crustal layers, $P_gP$ = intra-crustal reflection, $P_mP$ = Moho reflection, and $P_n$ = upper-most mantle refracted arrivals. e) Velocity model obtained after forward layered modelling, using Rayinvr. Color codes are the same than in c). The black lines present the model interfaces. See text for further explanations.*

Figure 8 shows an example of data, picking and the output models for Tomo2D and Rayinvr along the same profile, for comparison. The tomography model (Figure 8c) was published 5 years prior to the forward layered model (Figure 8e). It consists in preliminary results for the seismic refraction data. However, the authors did not have access to any coincident multi-channel deep seismic data and assumed a very simple starting model. Furthermore, this first processing did not allow to observe some very weak $P_n$ arrivals (yellow picks in Figure 8d). The layered forward model presents the final results of this study. The authors had access to high quality coincident multichannel deep seismic data, showing some insights on the sediment layers, basement morphology, as well as clear mid-crustal and Moho reflectivity. To a first order, the two models show similar velocities and Moho depths. However, there are some differences that should be explained. The Moho depths are shallower in the tomography model than in the forward layered model in the south-easternmost part of the model. The picks used as $P_mP$ reflections for the tomography were probably arrivals from the strong mid-crustal reflection rather than the Moho. The velocity discrepancies between the two models are due to the lack of knowledge of the shallow structures as well as some smearing. For example, the velocities of the tomography model at model distance of 80 km are too high compared to the forward layered model. Tomography will compensate by using velocities that are slightly too low underneath, allowing to fit the picked travel-times. Thus, this comparison shows the importance of data complementarity to help the modelling.

- Resolution assessment

Before moving to the interpretation step, it is important to assess the resolution of the velocity model. The first two mandatory pieces of information that should be found along each velocity model are the modelling statistics and the fits (or travel-time residuals). The modelling statistics consist in the presentation of number of rays traced (N), normalized misfit ($\chi^2$, average of the travel-time misfits to picking uncertainty ratios) and root-mean squared residual travel-time ($t_{RMS}$) for each seismic phase, or, at least the whole model. The fit figures show an overlay of the picks and computed arrival-times in the model, which should be as close as possible.

Another important thing to observe is where there is ray coverage in the model. This is usually shown as brighter areas in the model, or as a separate plot showing the seismic rays or the derivative weight sum (DWS, sort of measure of the ray density). Furthermore, areas where rays are traveling in only one direction (i.e., ends of the model) may be affected by non-uniqueness issues, and lower resolution.

Further resolution assessment methods allow to identify the resolution of an interface depth, or velocity values, but they are not always shown as the computing time may be very large. Indeed, they are based on realizing a large number of tests, or computing a "new" dataset:

a) Monte-Carlo simulations: The model and picks are randomized before running an inversion. A large number of simulations is necessary to estimate the statistical resolution of the model. The output shows standard deviations of the velocity field, and model interfaces (if applicable).
b) Checkerboard tests (tomography): Travel-times are computed in a perturbed model, following a checkerboard pattern before running a new inversion with these modified travel-times. The result of the inversion is compared to the velocity model, and the areas where the pattern are found are the areas where a velocity anomaly of similar size is well resolved. Many checkerboard patterns can be tested.
c) Sensitivity tests (forward modelling): This allows to look at the variations of model misfit with systematic changes in chosen model parameters (Moho depth, crustal velocity gradient, upper-most mantle velocities, etc.). The results are usually shown as a plot a model statistics (i.e. $\chi^2$ or $t_{RMS}$) together with the number of rays traced.
d) Synthetic seismograms (forward modelling): A synthetic wavefield can be computed from the velocity field and impedance contrasts at the interfaces. Synthetic seismograms may be used to justify the choice of a specific velocity gradient, or a feature that is not shown by the travel-times.

- Interpretation

Once the velocity model is finalized, the geological interpretation can be achieved in the light of the resolution assessment. The different velocities and bodies highlighted by the model may be interpreted in terms of geological structures, using the seismic velocities and structure morphology. Typical P-wave velocity values are listed below. However, note that any deformation, alteration, hydrothermal and/or geochemical process can lead to profound modifications of these values (see chapter XXX).

- the water (~1.5 km/s),
- sediments (1.6-4.5 km/s),
- continental crust (4.5-7 km/s, with a mid-crust at ~6.5 km/s),
- oceanic crust (oceanic layering with velocities ranging from 4.5 to 7+ km/s, and lower velocity gradient in the gabbros),
- unaltered mantle (7.8-8.3 km/s),
- partially serpentinised mantle (5-8 km/s depending on the degree of serpentinisation).

Anomalous velocity bodies are often observed along rifted margins, usually at the crust-mantle boundary. The velocity gradient and eventual reflection observed at the base of the body can help determine the nature of these features (see part B.1.c)).

## 2. Potential field methods

Oceanographic vessels are usually equipped with onboard gravimeter and towed magnetometer. The data acquired using these instruments gives information on the sub-seafloor geology using the density (gravity data) and/or the magnetization (magnetic anomaly data) variations.

### a) Gravity anomalies

•Acquisition

The gravimeter is installed near the center of motion of the ship, to minimize its vertical accelerations and is mounted on a stabilized platform to reduce the effect of the horizontal accelerations. It measures changes in the vertical component of the gravity field through relative gravity measurements, i.e. the gravimeter is not absolute. Relative gravity data needs to be tied to a reference data, i.e. absolute measures of the gravity field (in harbors), in order to also correct for the system drift. The accuracy of modern marine gravity data is generally less than 1 mGal, thanks to enhancement of gravimeter and navigational systems. Moreover, those datasets are completed by satellite-derived gravity data (using satellite altimetry) and airborne data. These latter are less accurate but have a uniform coverage.

•Data processing

In marine geophysics, the free-air gravity anomaly is computed from raw data. The first step is to remove and filter the strong vertical anomalies due to the accelerations of the ship. The acceleration due to the motion of the ship at the Earth's surface is also corrected (Eotvos correction). The free-air gravity anomaly results from the subtraction of the normal gravity field of the Earth (gravity effect of the reference ellipsoid taking into account the Earth's rotation) from the recorded gravity data. The residual gravity, or Bouguer anomaly, is obtained by subtracting the gravity contrast produced by the water–sediment interface and the sediment-crust interface from the Free-air anomaly. Hence, the obtained signal reflects variations in the thickness of the crust or lateral variations of density on the crust.

•Modelling and interpretation

When considering gravity anomalies, it is important to remember that the shape and amplitude of these anomalies are highly dependent on the distance between the measurement location and the considered object. The farther the object, the wider the anomaly and the lower the amplitude (Figure 9).

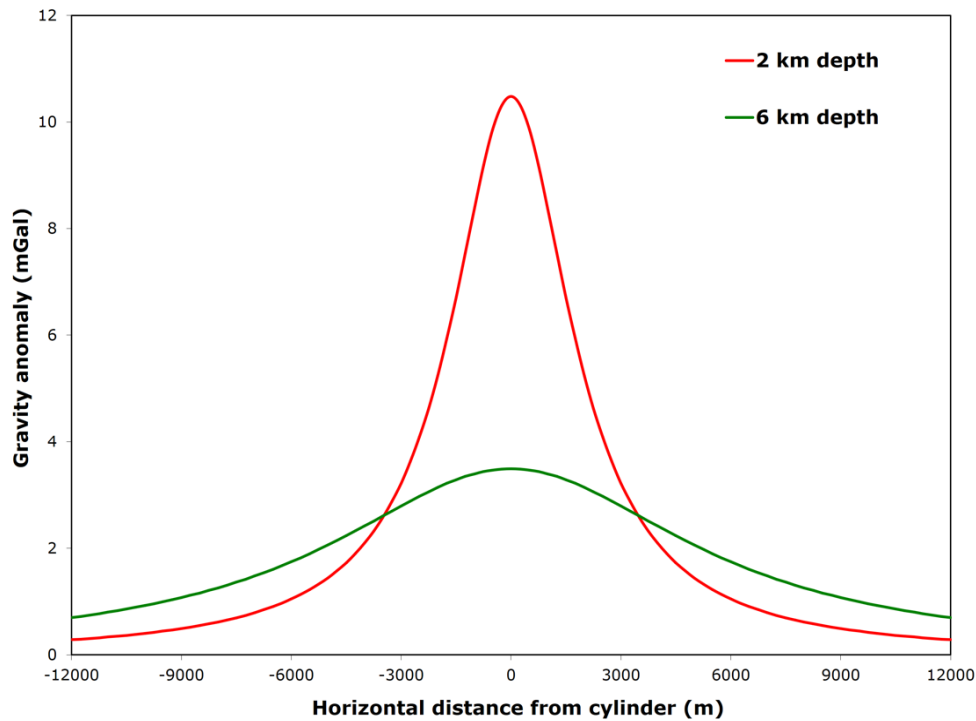

*Figure 9: Gravity anomaly for 2 small cylinders at 2 and 6 km depth. Each cylinder has the same diameter (2 km) and the same density contrast with the surrounding rocks (500 kg.m$^{-3}$).*

Free-air gravity anomalies are often used to perform forward gravity modelling (Figure 10). A simple structural model is built from seismic information (depth-converted multi-channel seismic profiles and/or velocity models, converted into densities using empirical relationships). It contains blocks of constant densities that can be interpreted in terms of nature of material. The forward modelling computes the free-air anomaly generated by the structural model. This modelled anomaly is then compared to the measured one. A good fit of modelled and measured data is considered as an indication that the inferred model is possible. However, it is common knowledge that such models are non-unique. Indeed, different structural models can lead to a good fit of the data. It is particularly true if the modelling is achieved along a single seismic profile in 2D. Multiplication of modelled profiles along the same area, or, even better, 3D modelling may be considered as more reliable. Another approach is gravity inversion, i.e., computation of an interface depth, and/or of a layer density, to fit the measured data. Here again, the non-uniqueness of the data is a problem. Gravity models can be performed together with isostasy calculation, allowing to better constrain the models. However, there is a strong initial assumption that the area is in isostatic equilibrium.

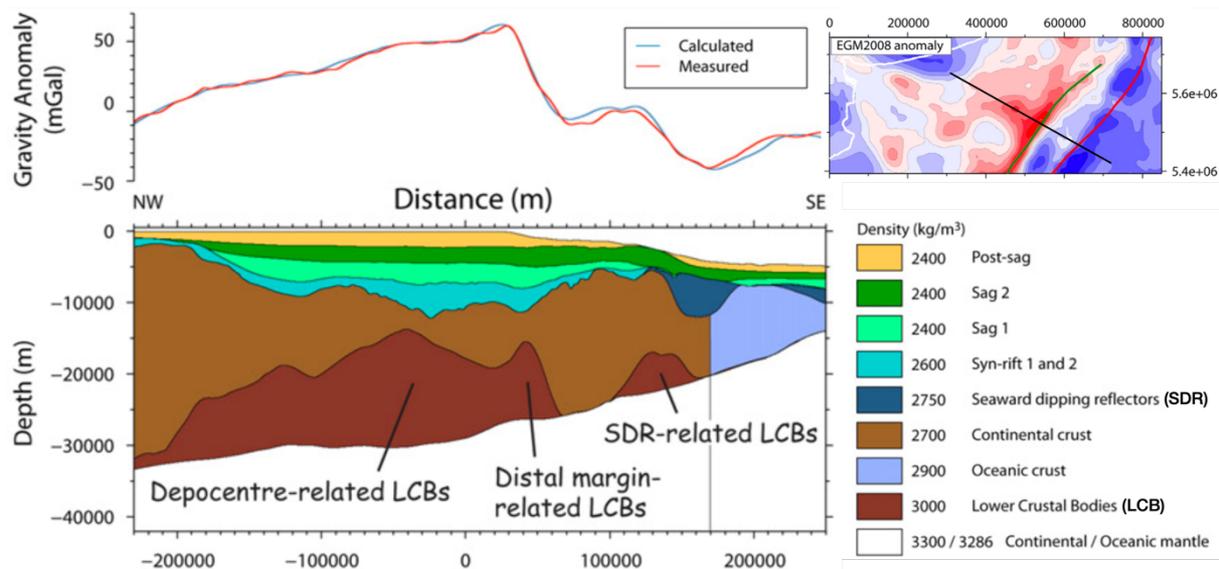

*Figure 10: Example of gravity modelling in the Colorado Basin, Atlantic Argentine margin (Autin et al. 2016). Each sediment layer or crustal body is assigned a single density value (bottom right). The geometry is known from seismic profiles, except for the lower crustal bodies (bottom left), which geometry is inferred from the gravity modelling (top left). Location of the profile on the gravity map (top right).*

The gravity signal varies strongly along rifted margins, from the proximal continental crust toward the denser oceanic crust (Figure 11). It can be used to localize the ocean-continent transition (OCT) but it is generally more reliable if it is corrected from the influence of the sedimentary thickness. Local high amplitudes in the OCT may be attributed to high-density bodies, related to either (1) underplated or intruded magmatic bodies at volcanic rifted margins, or (2) mantle exhumation, as the dense mantle (partly serpentinized) is located at shallower depth (<6 km) than below oceanic crust (6-8 km).

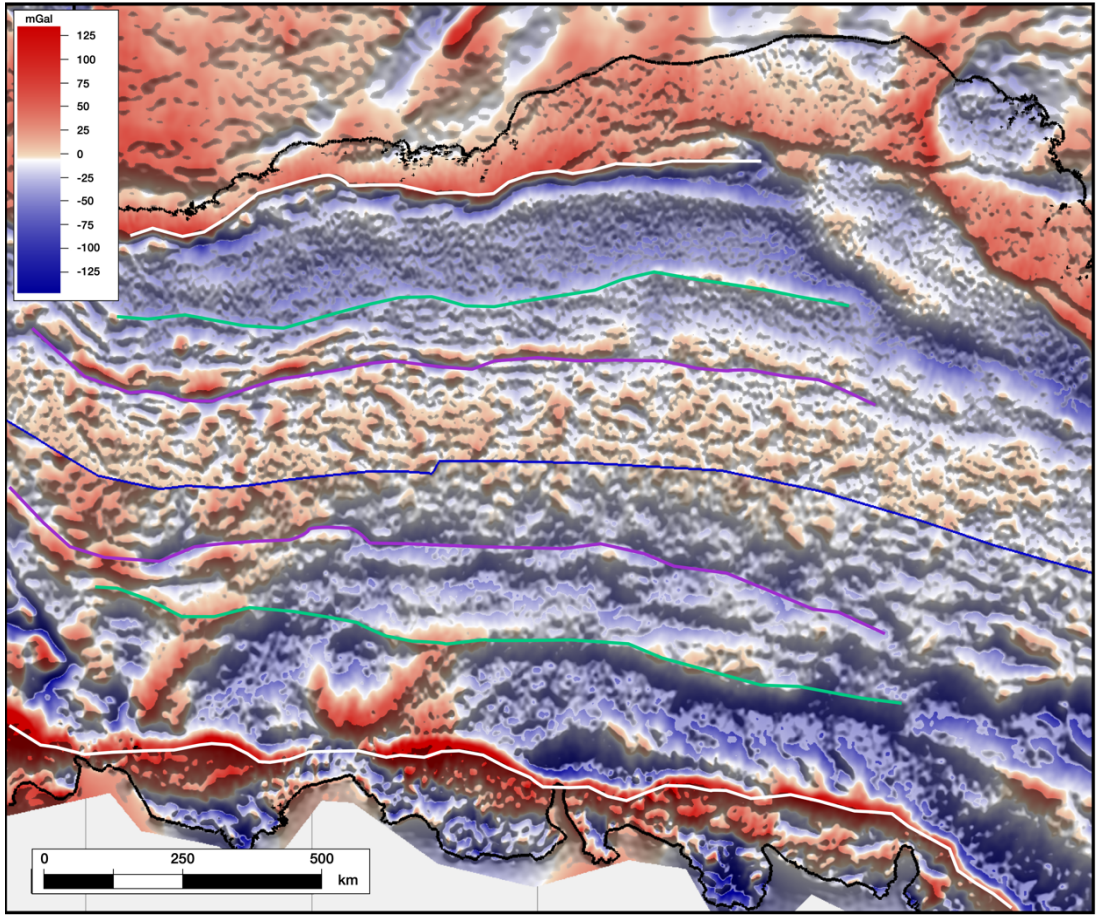

*Figure 11: Free-air gravity anomaly map of Sandwell and Smith (2009) for the conjugated Australia-Antarctica margins (restored to the C18 anomaly, blue line, 38 Ma). Projection: UTM50S. The shadow effect is the north-south horizontal gradient of free-air gravity anomalies derived from Sandwell and Smith [2009]. Boundaries from Gillard et al. (2015). White: continental crust necking. Green: limit between hyper-extended continental crust and exhumed mantle domain. Purple: limit between proto-oceanic crust and oceanic crust.*

### b) Magnetic anomalies

- Acquisition and processing

Intensity of the magnetic field is recorded using an absolute magnetometer towed behind the ship at a distance of at least three times the length of the ship. Magnetic anomalies result of the subtraction of the intensity of the International Geomagnetic Reference Field (IGRF) at the location of the measurement from the recorded data. The temporal variations of the geomagnetic field must be considered, generally using data of a magnetic observatory or using complementary measures of a gradient magnetometer. The measurements accuracy is generally of 1 nT (nanoTesla).

- Modelling and interpretation

The shape and amplitude of magnetic anomalies are highly dependent on the distance between the considered object and the measurement location. Similar to gravity anomalies, the farther the object, the wider the anomaly and the lesser the amplitude. Moreover, the anomaly shape is not only linked to the object geometry, it also depends on the magnetization and regional magnetic field directions. As a result, the signal of magnetic anomaly is always formed by a positive and a negative part. The magnetic anomalies mainly result from lateral contrasts of magnetization rather than the magnetization intensity. This magnetic intensity is often considered as a remanent magnetization, e.g., for a given rock, the magnetization acquired in the past that is proportional to the reference magnetic field that occurred when the rock cooled.

Also, similarly to gravity anomaly interpretations, forward modelling can be performed to explain magnetic anomalies. However, more parameters have to be taken into account. Indeed, the modelled anomaly depends on the structural model, on the magnetic susceptibility assigned to each object and on the magnetic field direction. Forward modelling computes the magnetic anomaly generated by the structural model and assigned magnetic parameters. A good fit between the modelled anomaly and the measured one is considered as an indication that the inferred model is possible. Again, such models are non-unique, i.e., different structural models/magnetic parameters can lead to a similar good fit of the data. Multiple 2D profiles or 3D modelling may help reduce this non-uniqueness. Another approach is magnetic inversion, i.e., calculation of one interface depth or one magnetic parameter of a layer to fit the measured data. Here again, the non-uniqueness of the model is a problem.

Along rifted margins, magnetic maps are powerful tools to understand the breakup history of the margins (Figure 12). Oceanic-crust-related magnetic anomalies record the Earth magnetic field reversal (EMFR), showing isochrones. However, at the OCT it is not clear whether linear magnetic anomalies are EMFR isochrones or not. Indeed, along volcanic rifted margins, the magma excess that characterizes the OCT could generate a linear magnetic anomaly in contrast with the poorly magnetized continental crust (e.g. the East Coast Magnetic Anomaly (North Atlantic Ocean), Biari *et al.* 2017). At magma-poor rifted margins, the same contrast-related anomaly could be generated by mantle exhumation (magnetized minerals formed by mantle serpentinization) juxtaposed to the poorly magnetized continental crust. In both cases the anomaly is not necessarily an isochrone.

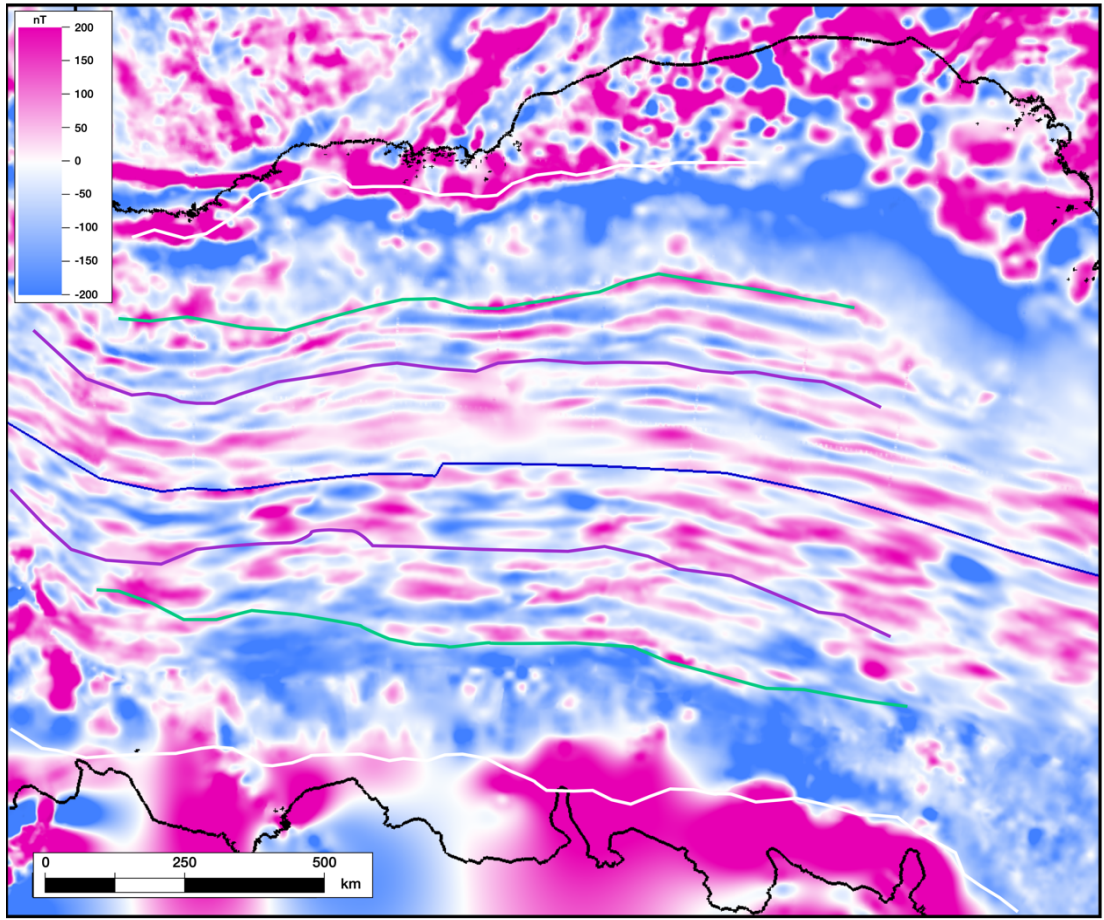

*Figure 12: Magnetic anomaly map from EMAG2 grid model (Maus et al. 2009) for the conjugated Australia-Antarctica margins (restored to the C18 anomaly, blue line, 38 Ma). Projection: UTM50S. Boundaries from Gillard et al. (2015). Color code of the lines is the same as figure 10.*

# B. Understanding continental rifted margins using geophysics

## 1. Geological objects interpretation

There is a large spectrum of interpretations for rifted margins. Sometimes it is linked to the large diversity of rifted margins, sometimes it is linked to different geological scenarios. It is thus important to understand how each geological or geophysical object is recognized and their possible interpretation. Figure 13 represents sketches of the two end-member rifted margins: magma-rich or magma-poor rifted margins. Each geological object is described in the following sections and figures.

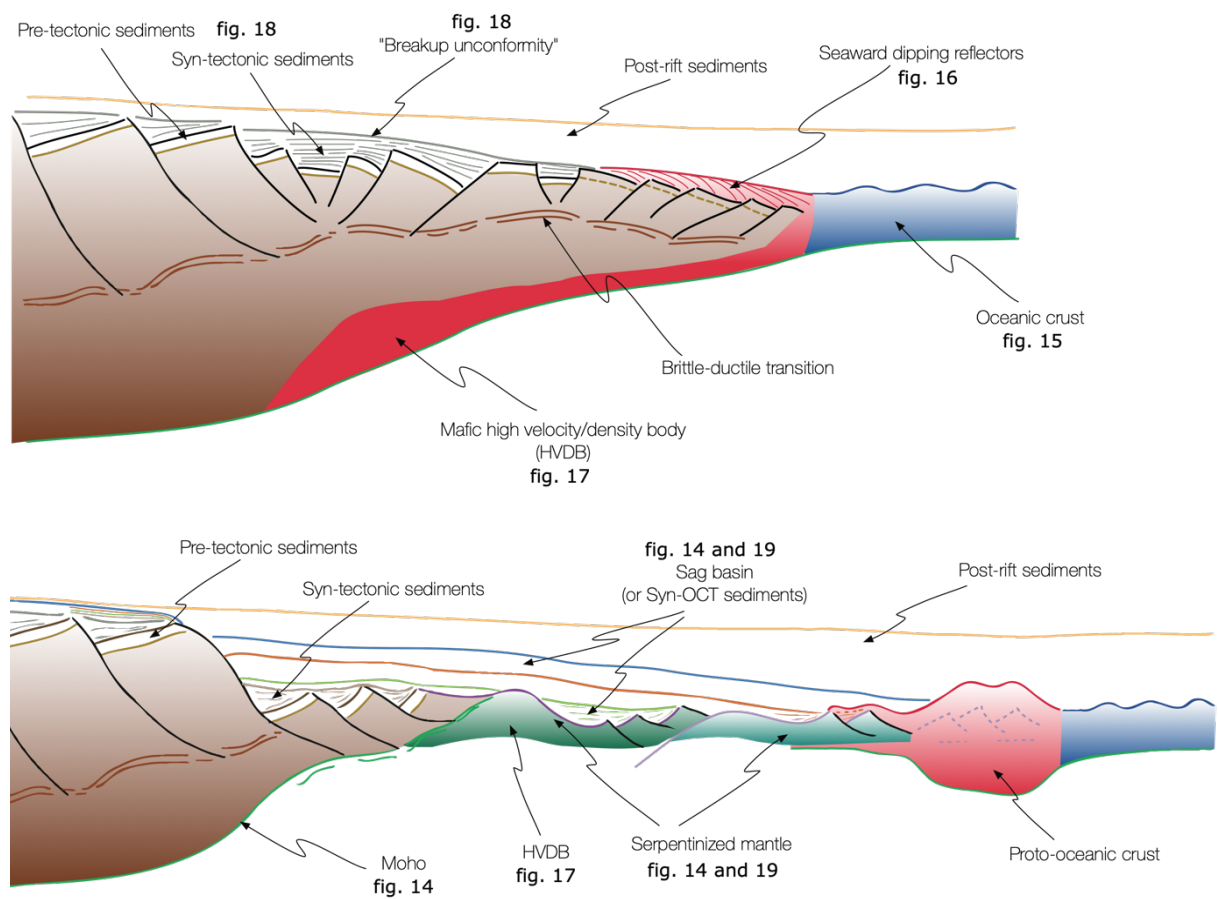

*Figure 13: Sketches of the two end-member rifted margins: magma-rich rifted margins (top) and magma-poor rifted margins (base). Each geological object is illustrated by different methods in the indicated figures. "Proto-oceanic crust": basement affected by large magmatic additions before the emplacement of steady-state oceanic crust.*

### a) Moho interface

The Moho is a petrological interface between mantle rocks and crustal rocks. When observed through geophysical methods, the Moho is mainly observed through the supposed density contrast between mantle rocks (3,3) and crustal rocks (2,7 to 2,9 kg/cm$^3$). Indeed, gravity models are built considering such density contrasts (modelled gravity Moho). P-wave velocities from seismic refraction modelling are dependent on the densities and lithologies (seismic velocity Moho), as well as reflectivity through the impedance contrast in seismic reflection and seismic refraction (reflective Moho or seismic Moho). It is thus imaged clearly and with high confidence below the continental crust (Figure 14) and the magmatic oceanic crust (Figure 15). However, in the transitional domain, the seismic velocity Moho (8 km/s) is always present, although the petrological Moho may be absent (e.g., below exhumed mantle domains at magma-poor margins). It can also correspond to other interfaces (fresh mantle/ serpentinized mantle, mantle/ "hot" material in the case of recent underplated material (Watremez *et al.* 2011)). The Moho reflection is present when the impedance contrast is strong enough. Thus, it is absent below exhumed mantle domains as the density gradient is very progressive, following the serpentinization degree (Figure 14).

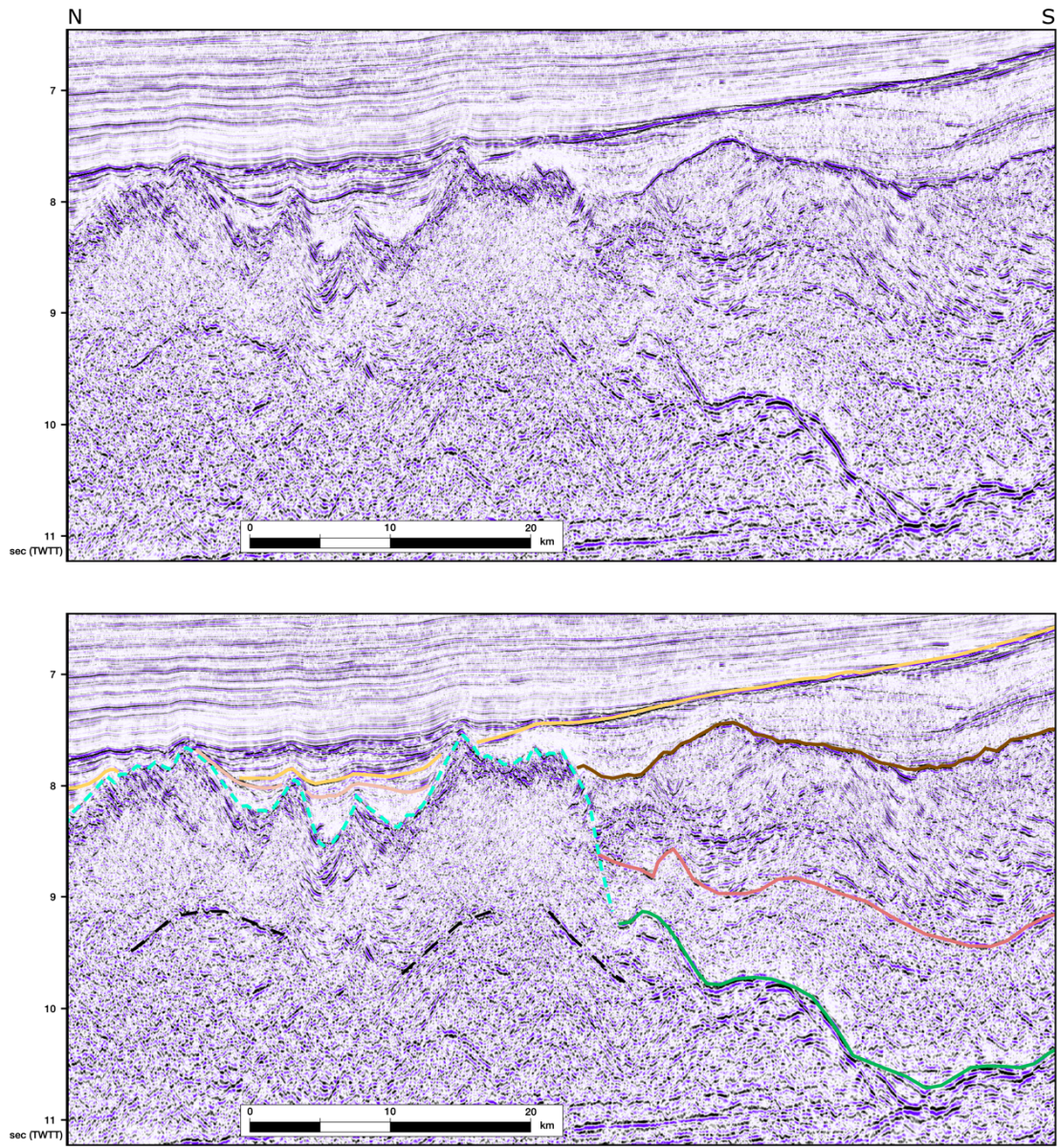

*Figure 14: Seismic reflection profile GA228-24 through the transition between continental crust (right) and exhumed mantle (left) without (top) and with interpretation (bottom, after Gillard et al. 2015). Green line: continental Moho. Pink line: top of the continental crust. Brown line: top of the pre-exhumation sediments. Yellow line: Top of the sag basin, i.e. oceanic breakup unconformity. Blue dashed line: top of the exhumed mantle. Black dashed line: it is not the Moho but a decollement level where normal faults root into, possibly controlled by the serpentinization degree of the exhumed mantle (Gillard et al. 2019).*

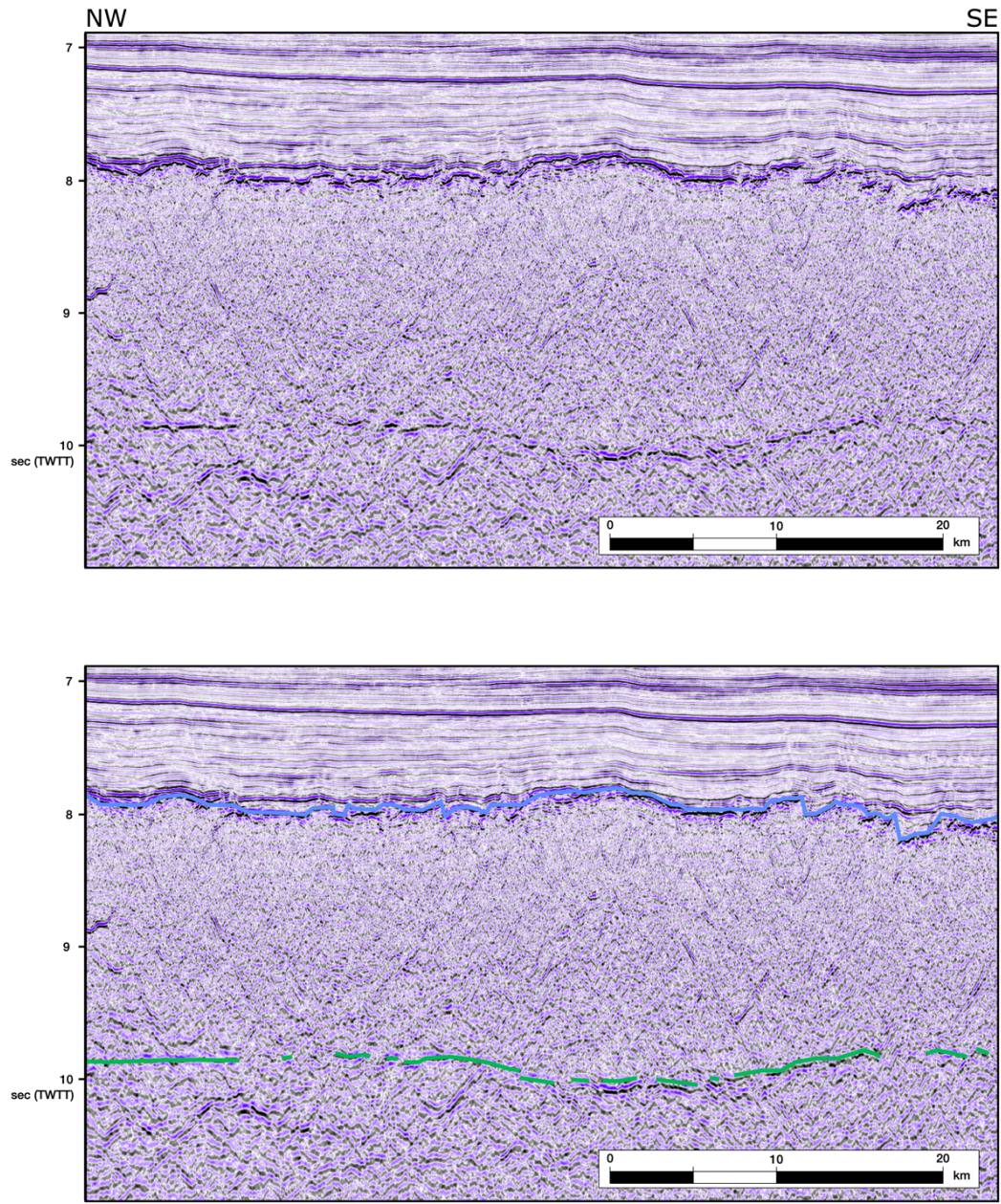

*Figure 15: Seismic reflection profile GA229-22 in oceanic crust without (top) and with interpretation (bottom). Green line: oceanic Moho. Blue line: top of the oceanic crust.*

### b) Magmatic additions

Magmatic objects are frequently observed at rifted margins, in particular, at magma-rich margins. Indeed, the first criteria used to recognize magma-rich margins is the presence of SDR (Seaward Dipping Reflectors, Figure 16), that are specifically observed on seismic reflection profiles. They are defined as wedges of stacked, arcuate reflectors that dip seawards with increasing flow thickness down dip (e.g. Mutter *et al.* 1982; Harkin *et al.* 2020). Geologically, they are interpreted as lava flows, interbedded with sediments, that characterized the high magmatic budget of the magma-rich margins, during the breakup and the formation of a steady-state mid-oceanic ridge (see chapter XXX). Another typical feature observed with deep seismic reflection and/or seismic refraction is an intruded lower crust (see the following paragraph B.1.c for further details).

Other magmatic objects can be recognized as sills, dikes, volcanic seamounts or lava flows and deltas (see chapter XXX). In seismic reflection profiles, magmatic additions generate strong amplitudes. Sills often appear clearly in the sedimentary succession, with very high amplitude that can stop abruptly laterally. Volcanic seamounts are frequent and of various sizes. They are dome shaped and generally show no sharp edges (Figure 16). When associated with normal faults they change the faulted block geometry: instead of being asymmetric, they tend to be more symmetric and dome shaped.

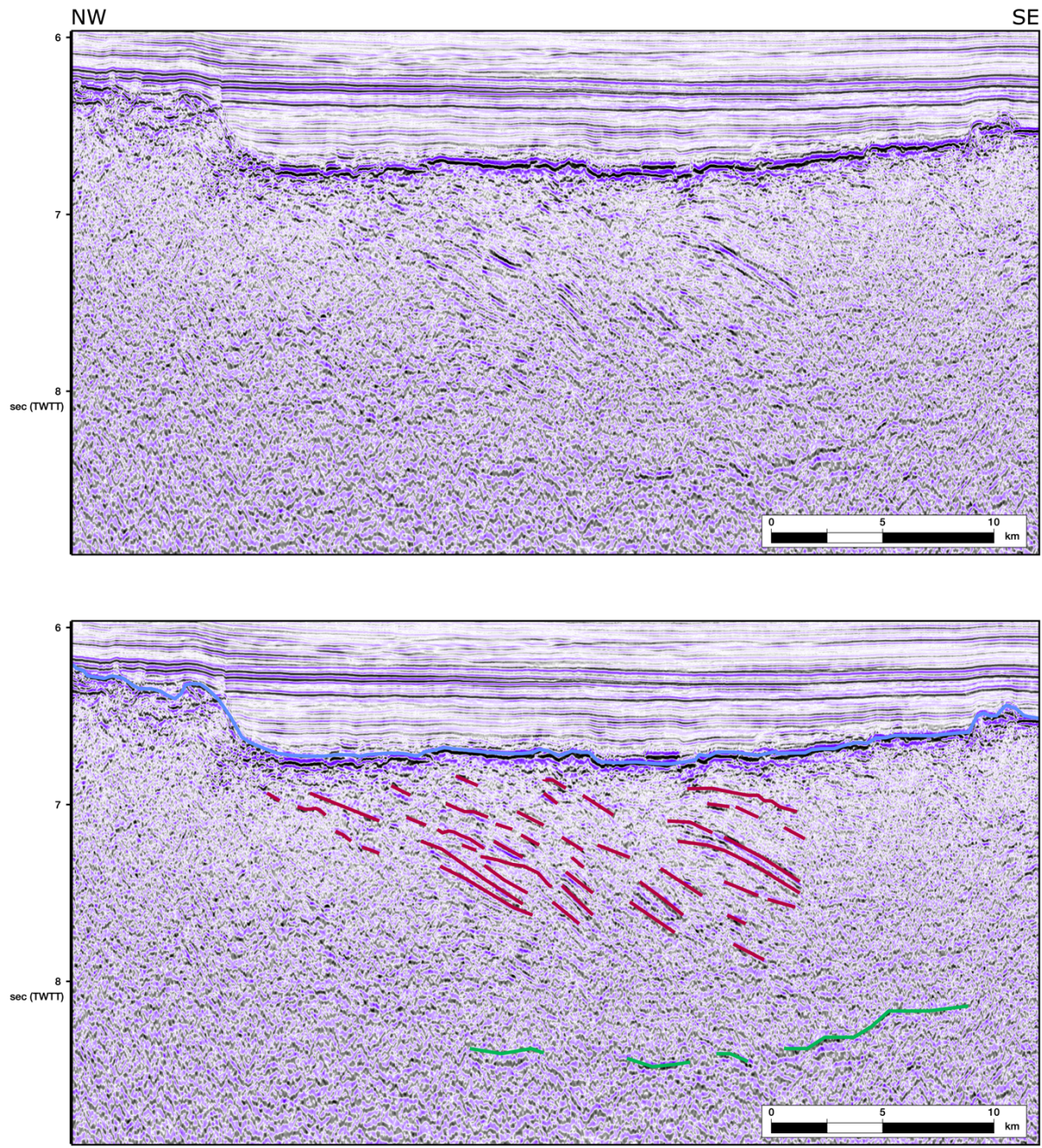

*Figure 16: Seismic reflection profile GA228-10 through seaward dipping reflectors (SDR) without (top) and with interpretation (bottom). Green line: transitional Moho. Blue line: top of crust. Red lines: SDR. Please note the volcanic seamount, as an acoustic basement high, on the left of the profile.*

### c) High velocity/density bodies (HVDB)

Rifted margins often show the presence of high velocity/density bodies (HVDB) in or at the base of the "crust". They are recognized through seismic refraction (with $V_P > 7.1$ km/s and $V_P < 8$ km/s) or in the gravity signal or modelling. They are also called "lower crustal bodies" (LCB) because of their location at the base of the continental crust or in the OCT deep basement, or "intermediate bodies" as their velocity and density characteristics are higher than typical crust and lower than unaltered mantle. Several hypotheses can explain their lithology and emplacement (Figure 17):

> (1) mafic intrusions emplaced in the lower crust during the rifting and often related to Seaward Dipping Reflectors (SDRs) at the surface (e.g. Eldholm *et al.* 2000; White, McKenzie 1989),
> (2) serpentinized mantle exhumed during the final extreme thinning of the crust (e.g. Gernigon *et al.* 2004; Boillot *et al.* 1987), or
> (3) high degree metamorphic rocks inherited from a previous orogenic phase (e.g. Gernigon *et al.* 2004; Ebbing *et al.* 2006).

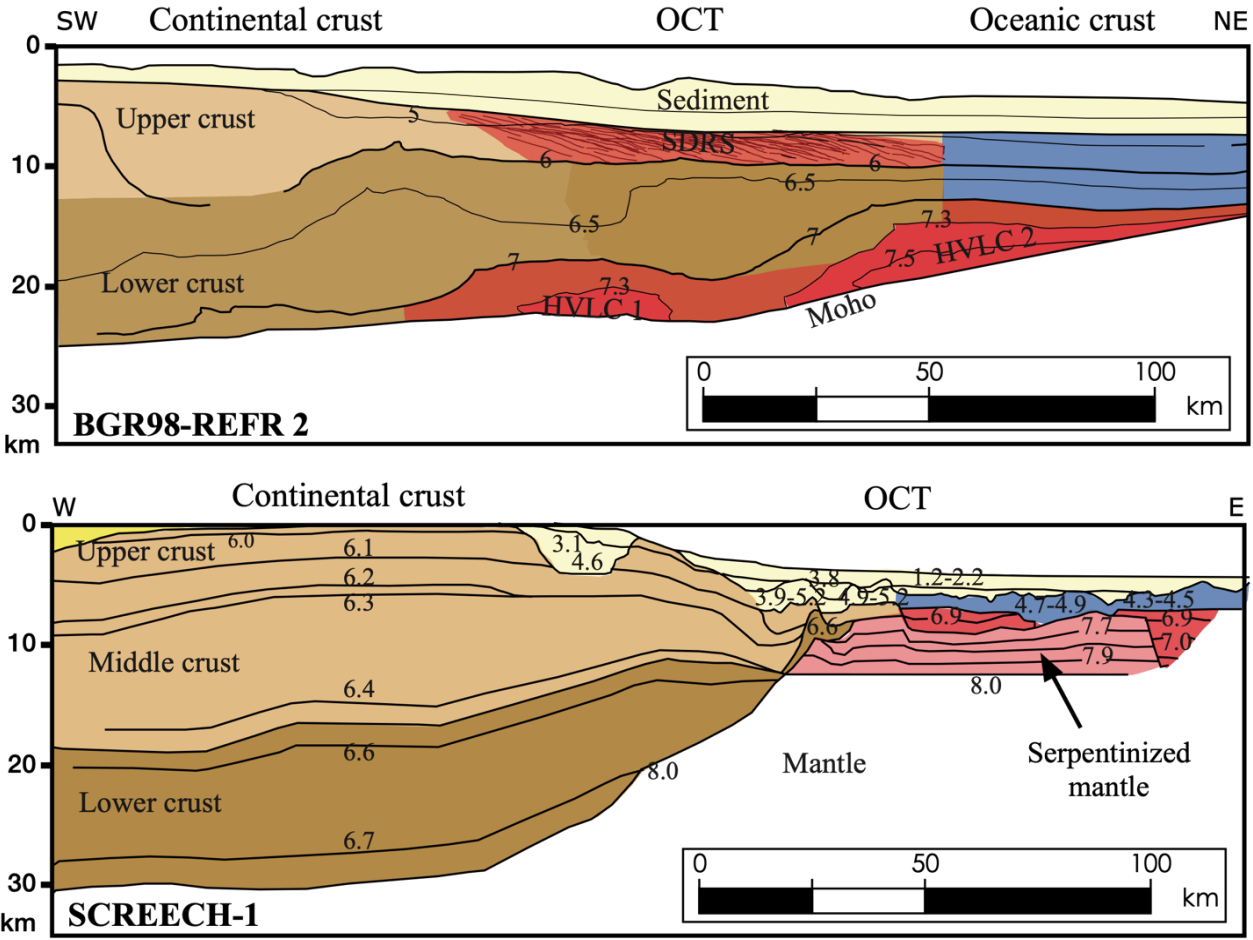

*Figure 17:* Deep structure of the Argentinian (top) and Newfoundland (bottom) continental margins (from Biari et al. 2021; modified after Schnabel et al. 2008; and Funck et al. 2003). The magma-rich Argentinian margin displays SDR and deep HVDBs (here HVLC for High Velocity Lower Crust) that are interpreted as mafic intrusions. The magma-poor Newfoundland margin displays a HVDB that is interpreted as serpentinized mantle.

It is a difficult task to determine the geometry and lithology of a HVDB, but it is of first importance to understand their role in the development of rifted margins. To discriminate between mafic intrusions and exhumed mantle, the vertical velocity gradient may be relevant:

gradients higher than 0.08 s$^{-1}$ are interpreted as typical of serpentinite, while gradients lower than 0.05 s$^{-1}$ may characterize mafic body (Watremez *et al.* 2011). Moreover, mafic bodies correlate with greater thicknesses (8-18 km versus 3-5 km for serpentinite) (Watremez *et al.* 2011). Another method consists in considering the thermal properties of such material (thermal conductivity and radiogenic heat production). It allows to calculate the conductive thermal field and to compare with available measured temperatures (Autin *et al.* 2016). Typical thermal properties are listed in Table 1. If mafic intrusions and inherited metamorphic rocks show only slight differences, serpentinized mantle has much lower thermal conductivity and heat production values. These values are derived from rock property records (e.g. Landolt, Börnstein 1983; Vilà *et al.* 2010). The conductivity of serpentinite is considered as half that of typical mantle minerals (White *et al.* 2003).

| Material composing HVDB | Bulk thermal conductivity (mW.K$^{-1}$) | Heat production (µW.m$^{-3}$) |
|---|---|---|
| mafic intrusions | 3,2 | 0.33 |
| serpentinized mantle | 2 | 0.007 |
| inherited high degree metamorphic rocks | 3,1 | 0.25 |

*Table 1:Thermal properties for the possible material composing HVDB.*

Of course, at well-known magma-rich margins, such bodies are generally interpreted as intruded magmatic/mafic bodies in the crust, or as underplated and cooled magma below the crust. The difference in the interpretation is important, as the location of the Moho depends on it: either the crust was thickened by intrusions, but the Moho still remain at the base of the HVDB, or the crust was underplated and the paleo-Moho is located at the top of the HVDB. Here, no discriminant method is available.

### d) Sedimentary units

Originally, the sedimentary seismic packages were systematically referred to as pre-rift, syn-rift and post-rift. However, when describing seismic units, the observer should not use such interpretative terms at first. The observation allows to describe pre-, syn- or post-tectonic units (Figure 18). All syn-tectonic units are not coeval and post-tectonic units can either be syn-rift or post-rift (Figure 19, see chapter XXX). In other words, syn-rift sediments can have various shapes and organization that also depends on the type of fault affecting them: classical normal faults or detachment faults. Ideally, the observer is able to recognize unique seismic facies for each sedimentary unit. This allows to correlate sediments strata of the same age with different tectonic patterns: for example, distal syn-tectonic strata with more proximal post-tectonic ones (Figure 19). Such observations highlight the migration of the deformation through the rift and its contribution to subsidence. However, as the sedimentary setting is prone to change laterally from proximal to distal, the seismic facies should change as well. The proximal/distal correlation is thus often difficult and based on strong assumptions.

The following descriptions highlight the main characteristics of pre-syn- and post-tectonic units deposited along a classical normal fault (Figure 18).

- Pre-tectonic units are not always easily observed as they generally present a conform relationship with the underlying top basement, i.e. the reflectors are parallel to the basement top. It is thus sometimes difficult to differentiate between pre-tectonic sediments and a layered top basement. They are characterized by tilted and parallel reflectors above the tilted blocks in between normal faults. They present toplap terminations, upward, and downlap terminations, downward, where the unit is cut by normal faults.
- Syn-tectonic units present typical wedge-shaped organization. The basal reflectors present a conform relationship with the top of the pre-tectonic units. The top reflectors are often sub-horizontal. In between, the reflectors are progressively decreasing their tilted angle to form a sedimentary wedge. The reflector terminations are either downlaps or onlaps against the bounding normal fault.
- Post-tectonic units are characterized by parallel reflectors that can either be sub-horizontal or tilted toward the ocean. The basal reflectors often present an unconformity with the top of the syn-tectonic units. The reflector terminations are either onlaps or downlaps above the underlying units.

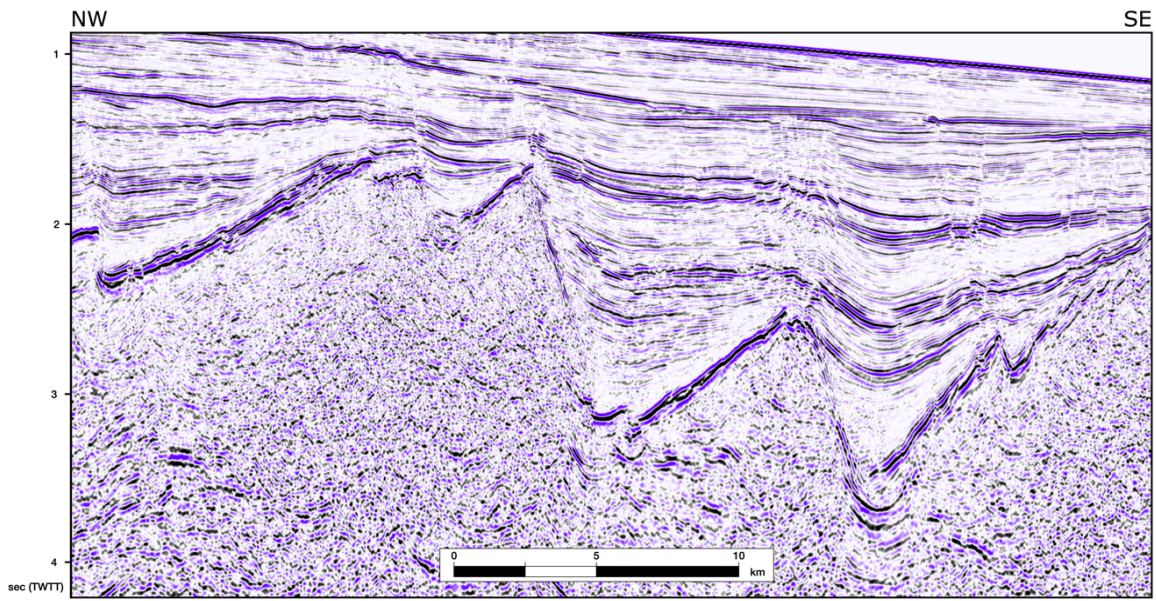

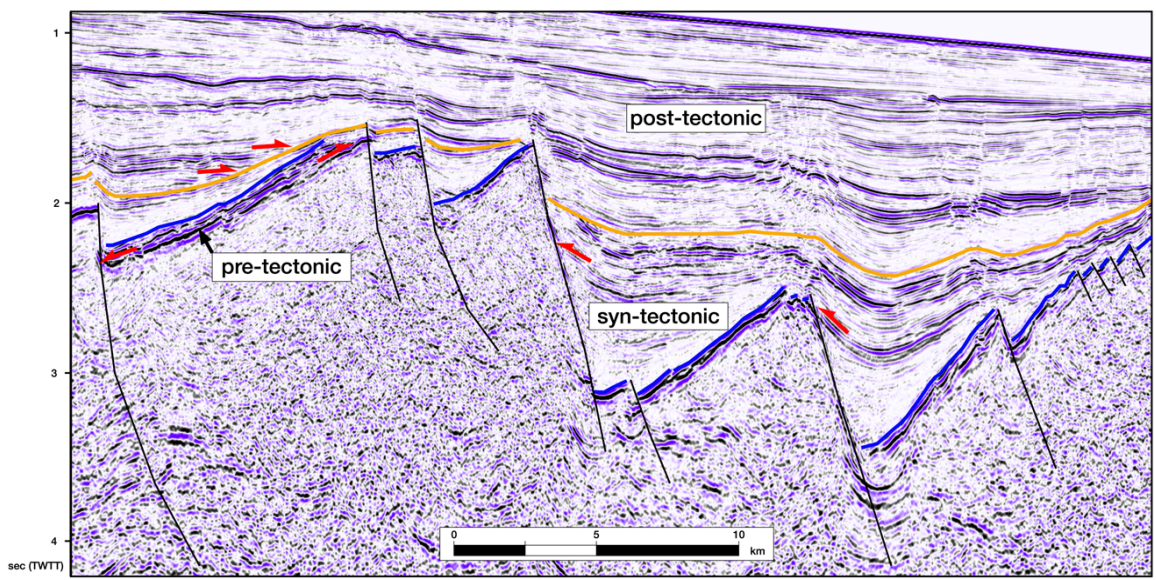

*Figure 18: Seismic reflection profile GA199-09 through the proximal margin without (top) and with interpretation (bottom). Blue line: top of pre-tectonic sediments. Orange line: top of syn-tectonic sediments and base of the post-tectonic sediments that onlap above the unconformity. Red arrows highlight the termination of reflectors in each unit (see Figure 6).*

### e) Unconformities

Another point of concern is the so-called "breakup unconformity" (see discussion in chapter XXX). It is defined as the interface separating the syn-rift and post-rift strata and is thought to date the breakup as an isochrone timeline. Yet, looking at interpreted breakup unconformities, it appears that it is often located at the top of syn-tectonic unit either in the proximal domain or in the distal domain. In such interpretations, it is most likely that the pointed unconformity corresponds to the end of different tectonic events and is thus not an isochrone timeline. Moreover, in its proximal part, it would not date the breakup but rather the end of the earlier syn-tectonic event (Figure 18 and Figure 19). Here, it is sometimes referred as the "necking unconformity". In its distal part, it could date the continental crust breakup and can be referred as the "crustal breakup unconformity". If there is a sag basin, the eventual unconformity at its top does date the breakup, i.e. the onset of spreading, and can be referred as the "oceanic breakup unconformity" (Figure 14 and Figure 19).

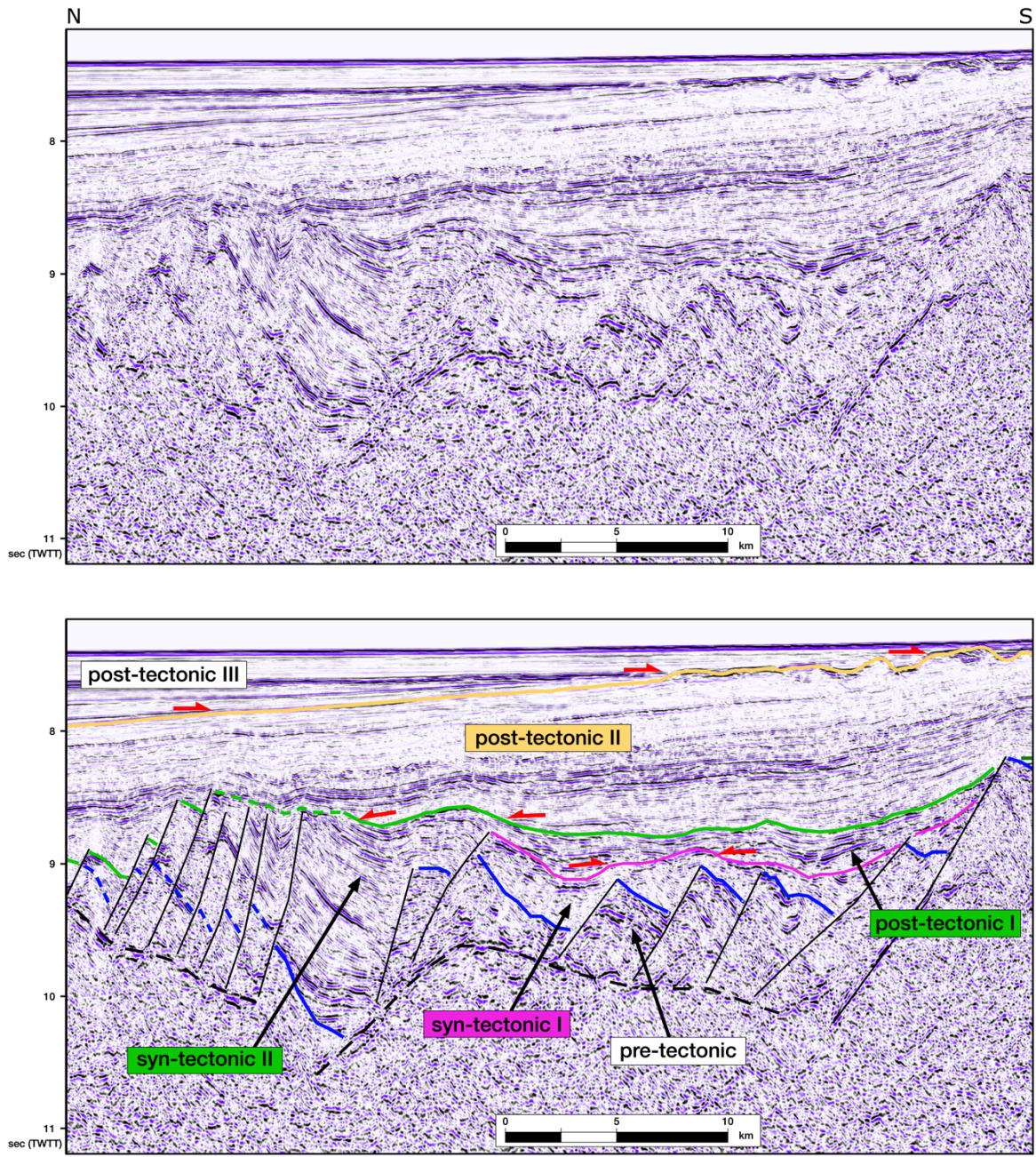

*Figure 19: Seismic reflection profile GA228-24 through the zone of exhumed mantle without (top) and with interpretation (bottom, after Gillard et al. 2015). Blue line: top of pre-tectonic sediments. Pink line: top of syn-tectonic I sediments and base of post-tectonic sediments that are coeval with the syn-tectonic II sediments. Green line: top of syn-tectonic II sediments and top of post-tectonic I sediments. Base of post-tectonic III sediments that onlap above the unconformity. Yellow line: Top of the sag basin and base of post-tectonic III sediments that onlap above the oceanic breakup unconformity. Black dashed line: it is not the Moho but a decollement level where normal faults root into, possibly controlled by the serpentinization degree of the exhumed mantle (Gillard et al. 2019). Red arrows highlight the termination of reflectors in each unit (see Figure 6).*

## *2. Data complementarity*

In the absence of rock samples, geophysical methods are the only way to study rifted margins. They provide indirect imaging of the crust and sediments as well as potential field data. Each method provides information of its own, but the understanding of the geological processes leading to the formation of the rifted margin is best with the combination of these complementary methods. Moreover, onshore/offshore approaches are also an additional bonus, as well as comparisons with field analogs.

To illustrate this powerful complementarity, the difficult location of the OCT boundaries in magma-poor-rifted margins is a good example. Geophysical characteristics of the OCT are neither typical of continental crust nor of oceanic crust. We illustrate this with Figure 20 along the North-Eastern margin of the Gulf of Aden where numerous experiments took place (Leroy *et al.* 2010; Watremez *et al.* 2011; D'Acremont *et al.* 2005; D'Acremont *et al.* 2006; Autin *et al.* 2010; Lucazeau *et al.* 2008; Leroy *et al.* 2012).

Looking at the magnetic signal, it shows two distinct domains of magnetic pattern: oceanward, the signal shows strong, continuous successive magnetic anomalies whereas, continentward, the signal is weaker, without clear oscillations. In the OCT, there are no clearly identifiable oceanic magnetic anomalies. This excludes that the OCT is composed of a magmatic oceanic crust and allows to locate the oceanward boundary of the OCT along anomaly 5d (D'Acremont *et al.* 2006). As for the wide-angle profile (Leroy *et al.* 2010), it shows a narrow zone of continental crust thinning, with thicknesses decreasing from 30 to less than 10 km thick. In the South, the oceanic crust has typical layers 2 and 3 velocities (5 to 5.5 and 6,5 to 7 km/s, respectively) and the thickness of a slow-spreading oceanic crust (ca. 6 km) (Christeson *et al.* 2019). In between, the 5-km-thick OCT shows a high velocity body at the base of the "crust". Such bodies could either result from a mafic body or from serpentinite. To discriminate between the two, vertical velocity gradient seems relevant: gradients higher than $0.08$ $s^{-1}$ are typical serpentinite, while gradients lower than $0.05$ $s^{-1}$ may characterize mafic body (Watremez *et al.* 2011). Moreover, mafic bodies correlate with greater thicknesses (8-18 km versus 3-5 km for serpentinite) (Watremez *et al.* 2011). Here the high velocity body is characterized with velocities rapidly decreasing upward from 7,25 km/s to 6,2 km/s with a gradient of $0,21$ $s^{-1}$ and a thickness of 5 km. This typical P-wave velocity increase at the base of the "crust" characterizes the presence of serpentinized mantle: it corresponds to increasing serpentinization degree upward and thus less dense material. Further evidence resides in the seismic reflection profile. First, there is no clear Moho reflections in the OCT, which indicate that no sharp velocity contrast allows for such reflection. This could again indicate a gradual downward decrease in the degree of mantle serpentinization. Then, if mantle exhumation occurs, the older sediments deposited on the continental crust, before the beginning of exhumation, are by definition restricted to the continental domain alone. Therefore, it is possible to locate the exhumation point (i.e. the northward boundary of the OCT) where the pre-exhumation sediments stop (Gillard *et al.* 2015). Moreover, the authors observe that a thick horizontal layer of sediments are deposited before the post-rift sediments (which are deposited above the oceanic crust) and after the sedimentary fans that are typical of the syn-rift period. Thus, these sediments were deposited during the continental break-up itself and are referred as "syn-OCT sediments" (D'Acremont *et al.* 2005; Autin *et al.* 2010) (defined as "sag sediments" in other margins). Indeed, the formation of the OCT, here by mantle exhumation along large-offset exhumation faults, does not create syn-kinematic sedimentary wedges but rather flat or slightly downlapping syn-kinematic units (Péron-Pinvidic *et al.* 2007; e.g. Gillard *et al.* 2015). It is interesting to highlight that the syn-OCT sediments are coeval with both mantle exhumation and on-land major uplift, associated with a phase of erosion, reflecting the general seaward tilting of the whole margin (Leroy *et al.* 2012).

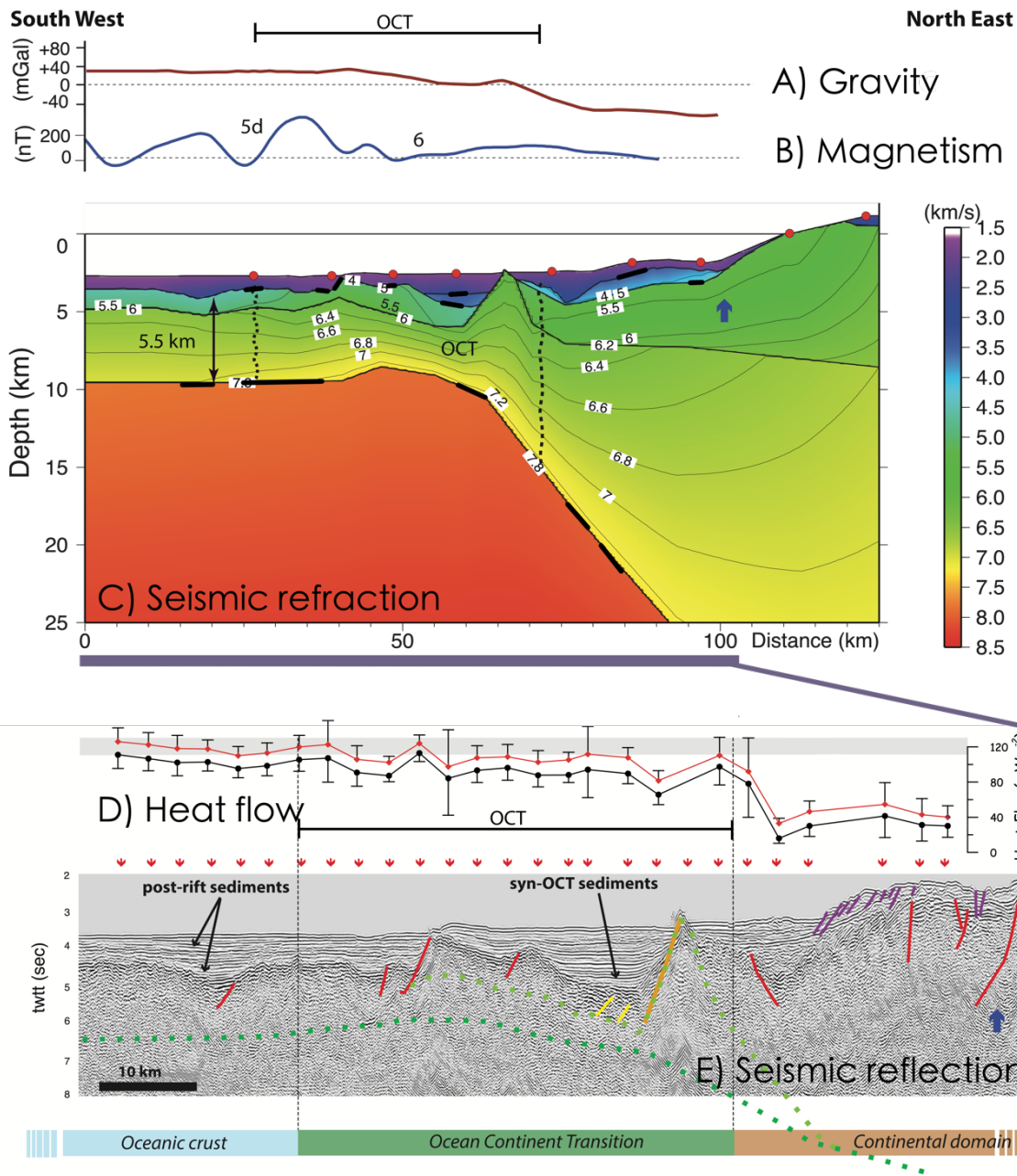

*Figure 20: Example of methods complementarity along the Mirbat segment in the North Eastern margin of the Gulf of Aden (Leroy et al. 2010; Leroy et al. 2012). A) Gravity signal is high above dense oceanic crust and low above continental crust. B) Above the oceanic crust, magnetic signal is strong with continuous successive magnetic anomalies oscillations whereas, continentward, the signal is weaker, without clear oscillations. C) The 5-km-thick OCT shows a high velocity body at the base of the "crust" with velocities rapidly decreasing upward characterizing the presence of serpentinized mantle. D) Heat flow is high above the conductive oceanic crust and low above continental crust. E) On the seismic reflection profile, faults active at the beginning of the syn-rift period are in red, faults active during the exhumation phase are in yellow, those active during the post-rift phase in orange and those associated with active submarine slides in violet. Colored dotted lines are derived from the velocity model, converted into two-way travel time; they correspond to the main interfaces underneath the acoustic basement.*

Finally, the presence of major deformation structures (large-offset normal fault) affecting both basement and overlying sediments are more typical of mantle exhumation processes than of oceanic crust formation (e.g. Gillard *et al.* 2015). These observations allow to locate the OCT between 30 and 70 km along the wide-angle profile: it must include the high velocity body as well as the syn-OCT sediments but no pre-exhumation sediments(Autin *et al.* 2010). Looking at the free-air gravity signal, it shows high positive amplitude in the oceanic domain (40 mGal). There are slightly higher amplitudes in the OCT above the basement highs, then a negative gradient toward strong negative continental amplitudes (-40 mGal). In the OCT, the high velocity body observed on wide-angle data, is associated with higher density than in the surrounding continental and oceanic crust. This lateral contrast can generate a positive gravity anomaly as observed above the interpreted serpentinized mantle high. The heat flow is also interesting as it shows high values in the conductive oceanic crust and in the OCT (120 mW.m$^{-2}$) and lower values in the less conductive continental crust (40 mW.m$^{-2}$). Such high values in the OCT are not common and may be explained by the fact that the Gulf of Aden is a young basin (less than 20 Ma). Thermal modelling (Lucazeau *et al.* 2008; Lucazeau *et al.* 2010) indicates that they could correspond to a thermal anomaly in the upper mantle which persisted after the break-up, possibly through small-scale convection cells below the margin during and after the rifting. If this behavior occurred in older rifted margins, it could explain the shallow water deposits during post rift-time.

## 3. Conclusion

Geophysical data are often the only constraints available for interpretation of offshore rifted continental margins. Seismic reflection, refraction and potential fields are all valuable datasets carrying relevant information. However, numerous uncertainties are inherent to the methods, which often hamper a proper validation of the geological model. The complementarity of the dataset as well as the large amount of data are favored to reach the most accurate interpretation. Nevertheless, even the most complete set of data can lead to various - if not contradictory - interpretation, due to the non-uniqueness of the models. The interpreter must convince his/her reader by gathering scientific proofs as well as giving arguments for a consistent geological scenario.

### *Acknowledgment*

(Gillard *et al.* 2017; Gholamrezaie *et al.* 2018; Szameitat *et al.* 2020)

## *References*

Autin, J., Leroy, S., Beslier, M.-O., d'Acremont, E., Razin, P., Ribodetti, A., Bellahsen, N., Robin, C., Al Toubi, K. (2010). Continental break-up history of a deep magma-poor margin based on seismic reflection data (northeastern Gulf of Aden margin, offshore Oman). [En ligne*Geophysical Journal International*], 180(2), 501–519. [En ligne]. Disponible à l'adresse : https://academic.oup.com/gji/article-lookup/doi/10.1111/j.1365-246X.2009.04424.x [Consulté le 12 April 2017].

Autin, J., Scheck-Wenderoth, M., Götze, H.-J., Reichert, C., Marchal, D. (2016). Deep structure of the Argentine margin inferred from 3D gravity and temperature modelling, Colorado Basin. [En ligne*Tectonophysics*], 676, 198–210. [En ligne]. Disponible à l'adresse : http://linkinghub.elsevier.com/retrieve/pii/S0040195115006526 [Consulté le 12 April 2017].

Biari, Y., Klingelhoefer, F., Sahabi, M., Funck, T., Benabdellouahed, M., Schnabel, M., Reichert, C., Gutscher, M.-A., Bronner, A., Austin, J.A. (2017). Opening of the central Atlantic Ocean: Implications for geometric rifting and asymmetric initial seafloor spreading after continental breakup: Opening of the Central Atlantic Ocean. [En ligne*Tectonics*], 26 June 2017. [En ligne]. Disponible à l'adresse : http://doi.wiley.com/10.1002/2017TC004596 [Consulté le 7 July 2017].

Biari, Y., Klingelhoefer, F., Franke, D., Funck, T., Loncke, L., Sibuet, J.-C., Basile, C., Austin, J.A., Rigoti, C.A., Sahabi, M., Benabdellouahed, M., Roest, W.R. (2021). Structure and evolution of the Atlantic passive margins: A review of existing rifting models from wide-angle seismic data and kinematic reconstruction. [En ligne*Marine and Petroleum Geology*], 126, 104898. [En ligne]. Disponible à l'adresse : https://www.sciencedirect.com/science/article/pii/S0264817221000027 [Consulté le 18 February 2021].

Boillot, G. *et al.* (1987). Tectonic denudation of the upper mantle along passive margins: a model based on drilling results (ODP leg 103, western Galicia margin, Spain). *Tectonophysics*, 132(4), 335–342. [En ligne]. Disponible à l'adresse : http://www.sciencedirect.com/science/article/B6V72-488953N-4H/2/857bb8b2715606703820d8036cf14afd.

Christeson, G.L., Goff, J.A., Reece, R.S. (2019). Synthesis of Oceanic Crustal Structure From Two-Dimensional Seismic Profiles. [En ligne*Reviews of Geophysics*], 57(2), 504–529. [En ligne]. Disponible à l'adresse : http://agupubs.onlinelibrary.wiley.com/doi/abs/10.1029/2019RG000641 [Consulté le 22 February 2021].

D'Acremont, E., Leroy, S., Beslier, M.-O., Bellahsen, N., Fournier, M., Robin, C., Maia, M., Gente, P. (2005). Structure and evolution of the eastern Gulf of Aden conjugate margins from seismic reflection data. [En ligne*Geophysical Journal International*], 160(3), 869–890. [En ligne]. Disponible à l'adresse : http://onlinelibrary.wiley.com.biblioplanets.gate.inist.fr/doi/10.1111/j.1365-246X.2005.02524.x/abstract [Consulté le 25 February 2013].

D'Acremont, E., Leroy, S., Maia, M., Patriat, P., Beslier, M.-O., Bellahsen, N., Fournier, M., Gente, P. (2006). Structure and evolution of the eastern Gulf of Aden: insights from magnetic and gravity data (Encens-Sheba MD117 cruise). [En ligne*Geophysical Journal International*], 165(3), 786–803. [En ligne]. Disponible à l'adresse : http://onlinelibrary.wiley.biblioplanets.gate.inist.fr/doi/10.1111/j.1365-246X.2006.02950.x/abstract [Consulté le 25 February 2013].

Dondurur, D. (2018). *Acquisition and Processing of Marine Seismic Data* Elsevier, Amsterdam.

Ebbing, J., Lundin, E., Olesen, O., Hansen, E.K. (2006). The mid-Norwegian margin: a discussion of crustal lineaments, mafic intrusions, and remnants of the Caledonian root by 3D density modelling and structural interpretation. *Journal of the Geological Society*, 163(1), 47–59. [En ligne]. Disponible à l'adresse : internal-pdf://Ebbing-al-2006-2106563072/Ebbing-al-2006.pdf.

Eldholm, O., Gladczenko, T.P., Skogseid, J., Planke, S. (2000). Atlantic volcanic margins: a comparative study. [En ligne*Geological Society, London, Special Publications*], 167(1), 411–428. [En ligne]. Disponible à l'adresse : http://sp.lyellcollection.org/content/167/1/411.abstract [Consulté le 20 December 2011].

Funck, T., Hopper, J.R., Larsen, H.C., Louden, K.E., Tucholke, B.E., Holbrook, W.S. (2003). Crustal structure of the ocean-continent transition at Flemish Cap: Seismic refraction resultst. *Journal of Geophysical Research*, 108(B11), 20 p. [En ligne]. Disponible à l'adresse : file://localhost/Volumes/Document/Aden/Biblio/Funck-al-2003.pdf.

Gernigon, L., Ringenbach, J.-C., Planke, S., Le Gall, B. (2004). Deep structures and breakup along volcanic rifted margins: insights from integrated studies along the outer Vøring Basin (Norway). *Marine and Petroleum Geology*, 21(3), 363–372. [En ligne]. Disponible à l'adresse : internal-pdf://Gernigon-al-2004-0445924096/Gernigon-al-2004.pdf.

Gholamrezaie, E., Scheck-Wenderoth, M., Sippel, J., Strecker, M.R. (2018). Variability of the geothermal gradient across two differently aged magma-rich continental rifted margins of the Atlantic Ocean: the Southwest African and the Norwegian margins. [En ligne*Solid Earth*], 9(1), 139–158. [En ligne]. Disponible à l'adresse : https://www.solid-earth.net/9/139/2018/ [Consulté le 22 February 2018].

Gillard, M., Sauter, D., Tugend, J., Tomasi, S., Epin, M.-E., Manatschal, G. (2017). Birth of an oceanic spreading center at a magma-poor rift system. [En ligne*Scientific Reports*], 7(1), 15072. [En ligne]. Disponible à l'adresse : http://www.nature.com/articles/s41598-017-15522-2 [Consulté le 3 September 2018].

Gillard, M., Autin, J., Manatschal, G., Sauter, D., Munschy, M., Schaming, M. (2015). Tectonomagmatic evolution of the final stages of rifting along the deep conjugate Australian-Antarctic magma-poor rifted margins: Constraints from seismic observations. [En ligne*Tectonics*], 34(4), 753–783. [En ligne]. Disponible à l'adresse : http://doi.wiley.com/10.1002/2015TC003850 [Consulté le 17 August 2018].

Gillard, M., Tugend, J., Müntener, O., Manatschal, G., Karner, G.D., Autin, J., Sauter, D., Figueredo, P.H., Ulrich, M. (2019). The role of serpentinization and magmatism in the formation of decoupling interfaces at magma-poor rifted margins. [En ligne*Earth-Science Reviews*], 196, 102882. [En ligne]. Disponible à l'adresse :

https://linkinghub.elsevier.com/retrieve/pii/S0012825218307013 [Consulté le 18 February 2021].

Harkin, C., Kusznir, N., Roberts, A., Manatschal, G., Horn, B. (2020). Origin, composition and relative timing of seaward dipping reflectors on the Pelotas rifted margin. [En ligne*Marine and Petroleum Geology*], 114, 104235. [En ligne]. Disponible à l'adresse : http://www.sciencedirect.com/science/article/pii/S0264817220300180 [Consulté le 15 December 2020].

Korenaga, J., Holbrook, W.S., Kent, G.M., Kelemen, P.B., Detrick, R.S., Larsen, H.-C., Hopper, J.R., Dahl-Jensen, T. (2000). Crustal structure of the southeast Greenland margin from joint refraction and reflection seismic tomography. [En ligne*Journal of Geophysical Research: Solid Earth*], 105(B9), 21591–21614. [En ligne]. Disponible à l'adresse : http://agupubs.onlinelibrary.wiley.com/doi/abs/10.1029/2000JB900188 [Consulté le 16 December 2020].

Landolt, H.H., Börnstein, R. (1983). *Landolt-Börnstein Zahlenwerte und Funktionen aus Naturwissenschaften und Technik: Neue Serie* Springer.

Leroy, S. *et al.* (2010). Contrasted styles of rifting in the eastern Gulf of Aden: A combined wide-angle, multichannel seismic, and heat flow survey. [En ligne*Geochemistry, Geophysics, Geosystems*], 11(7), n/a-n/a. [En ligne]. Disponible à l'adresse : http://doi.wiley.com/10.1029/2009GC002963 [Consulté le 12 April 2017].

Leroy, S. *et al.* (2012). From rifting to oceanic spreading in the Gulf of Aden: a synthesis. [En ligne*Arabian Journal of Geosciences*], 5(5), 859–901. [En ligne]. Disponible à l'adresse : http://link.springer.com/10.1007/s12517-011-0475-4 [Consulté le 12 April 2017].

Leroy, S., Cannat, M. (2014). MD 199 / SISMO-SMOOTH cruise,Marion Dufresne R/V. , 2014. [En ligne]. Disponible à l'adresse : https://campagnes.flotteoceanographique.fr/campagnes/14003300/ [Consulté le 18 February 2021].

Liang, Y., Delescluse, M., Qiu, Y., Pubellier, M., Chamot-Rooke, N., Wang, J., Nie, X., Watremez, L., Chang, S.-P., Pichot, T., Savva, D., Meresse, F. (2019). Décollements, Detachments, and Rafts in the Extended Crust of Dangerous Ground, South China Sea: The Role of Inherited Contacts. [En ligne*Tectonics*], 38(6), 1863–1883. [En ligne]. Disponible à l'adresse : http://agupubs.onlinelibrary.wiley.com/doi/abs/10.1029/2018TC005418 [Consulté le 16 December 2020].

Lucazeau, F., Leroy, S., Rolandone, F., d'Acremont, E., Watremez, L., Bonneville, A., Goutorbe, B., Düsünur, D. (2010). Heat-flow and hydrothermal circulation at the ocean–continent transition of the eastern gulf of Aden. [En ligne*Earth and Planetary Science Letters*], 295(3–4), 554–570. [En ligne]. Disponible à l'adresse : http://www.sciencedirect.com/science/article/pii/S0012821X10002852 [Consulté le 6 September 2012].

Lucazeau, F., Leroy, S., Bonneville, A., Goutorbe, B., Rolandone, F., d/'Acremont, E., Watremez, L., Dusunur, D., Tuchais, P., Huchon, P., Bellahsen, N., Al-Toubi, K. (2008). Persistent thermal activity at the Eastern Gulf of Aden after continental break-up. *Nature*

*Geosci*, 1(12), 854–858. [En ligne]. Disponible à l'adresse : http://dx.doi.org/10.1038/ngeo359 http://www.nature.com/ngeo/journal/v1/n12/suppinfo/ngeo359_S1.html.

Maus, S. *et al.* (2009). EMAG2: A 2–arc min resolution Earth Magnetic Anomaly Grid compiled from satellite, airborne, and marine magnetic measurements. [En ligne*Geochemistry, Geophysics, Geosystems*], 10(8). [En ligne]. Disponible à l'adresse : http://agupubs.onlinelibrary.wiley.com/doi/abs/10.1029/2009GC002471 [Consulté le 16 February 2021].

Mutter, J.C., Talwani, M., Stoffa, P.L. (1982). Origin of seaward-dipping reflectors in oceanic crust off the Norwegian margin by "subaerial sea-floor spreading". [En ligne*Geology*], 10(7), 353–357. [En ligne]. Disponible à l'adresse : https://doi.org/10.1130/0091-7613(1982)10<353:OOSRIO>2.0.CO;2 [Consulté le 15 December 2020].

Péron-Pinvidic, G., Manatschal, G., Minshull, T.A., Sawyer, D.S. (2007). Tectonosedimentary evolution of the deep Iberia-Newfoundland margins: Evidence for a complex breakup history. *Tectonics*, 26. [En ligne]. Disponible à l'adresse : file://localhost/Volumes/Document/Aden/Biblio/PeronPinvidic-al-2007.pdf.

Pichot, T., Delescluse, M., Chamot-Rooke, N., Pubellier, M., Qiu, Y., Meresse, F., Sun, G., Savva, D., Wong, K.P., Watremez, L., Auxiètre, J.-L. (2014). Deep crustal structure of the conjugate margins of the SW South China Sea from wide-angle refraction seismic data. [En ligne*Marine and Petroleum Geology*], 58, 627–643. [En ligne]. Disponible à l'adresse : http://www.sciencedirect.com/science/article/pii/S0264817213002602 [Consulté le 16 December 2020].

Sallarès, V., Gailler, A., Gutscher, M.-A., Graindorge, D., Bartolomé, R., Gràcia, E., Díaz, J., Dañobeitia, J.J., Zitellini, N. (2011). Seismic evidence for the presence of Jurassic oceanic crust in the central Gulf of Cadiz (SW Iberian margin). [En ligne*Earth and Planetary Science Letters*], 311(1), 112–123. [En ligne]. Disponible à l'adresse : http://www.sciencedirect.com/science/article/pii/S0012821X11005140 [Consulté le 16 December 2020].

Sandwell, D.T., Smith, W.H.F. (2009). Global marine gravity from retracked Geosat and ERS-1 altimetry: Ridge segmentation versus spreading rate. [En ligne*Journal of Geophysical Research: Solid Earth*], 114(B1). [En ligne]. Disponible à l'adresse : http://agupubs.onlinelibrary.wiley.com/doi/abs/10.1029/2008JB006008 [Consulté le 16 February 2021].

Schnabel, M., Franke, D., Engels, M., Hinz, K., Neben, S., Damm, V., Grassmann, S., Pelliza, H., Dos Santos, P.R. (2008). The structure of the lower crust at the Argentine continental margin, South Atlantic at 44°S. [En ligne*Tectonophysics*], 454(1–4), 14–22. [En ligne]. Disponible à l'adresse : http://www.sciencedirect.com/science/article/pii/S0040195108000759 [Consulté le 22 February 2013].

Szameitat, L.S.A., Manatschal, G., Nirrengarten, M., Ferreira, F.J.F., Heilbron, M. (2020). Magnetic characterization of the zigzag shaped J-anomaly: Implications for kinematics and breakup processes at the Iberia–Newfoundland margins. [En ligne*Terra Nova*], 32(5), 369–380. [En ligne]. Disponible à l'adresse : https://onlinelibrary.wiley.com/doi/10.1111/ter.12466 [Consulté le 19 November 2020].

Vilà, M., Fernández, M., Jiménez-Munt, I. (2010). Radiogenic heat production variability of some common lithological groups and its significance to lithospheric thermal modeling. [En ligne*Tectonophysics*], 490(3–4), 152–164. [En ligne]. Disponible à l'adresse : http://www.sciencedirect.com/science/article/pii/S0040195110002064 [Consulté le 22 February 2013].

Watremez, L., Leroy, S., Rouzo, S., d'Acremont, E., Unternehr, P., Ebinger, C., Lucazeau, F., Al-Lazki, A. (2011). The crustal structure of the north-eastern Gulf of Aden continental margin: insights from wide-angle seismic data. [En ligne*Geophysical Journal International*], 184(2), 575–594. [En ligne]. Disponible à l'adresse : http://onlinelibrary.wiley.com.biblioplanets.gate.inist.fr/doi/10.1111/j.1365-246X.2010.04881.x/abstract [Consulté le 25 February 2013].

Watremez, L., Helen Lau, K.W., Nedimović, M.R., Louden, K.E. (2015). Traveltime tomography of a dense wide-angle profile across Orphan Basin. [En ligne*GEOPHYSICS*], 80(3), B69–B82. [En ligne]. Disponible à l'adresse : http://library.seg.org/doi/10.1190/geo2014-0377.1 [Consulté le 22 February 2021].

White, N., Thompson, M., Barwise, T. (2003). Understanding the thermal evolution of deep-water continental margins. *Nature*, 426(6964), 334–343.

White, R., McKenzie, D. (1989). Magmatism at Rift Zones: The Generation of Volcanic Continental Margins and Flood Basalts. *J. Geophys. Res.*, 94(B6), 7685–7729. [En ligne]. Disponible à l'adresse : internal-pdf://White-McKenzie-1989-0920390145/White-McKenzie-1989.pdf.

Yilmaz, Ö. (2001). *Seismic Data Analysis*. Society of Exploration Geophysicists. [En ligne]. Disponible à l'adresse : https://library.seg.org/doi/book/10.1190/1.9781560801580 [Consulté le 22 February 2021].

Zelt, C.A. (1999). Modelling strategies and model assessment for wide-angle seismic traveltime data. [En ligne*Geophysical Journal International*], 139(1), 183–204. [En ligne]. Disponible à l'adresse : https://doi.org/10.1046/j.1365-246X.1999.00934.x [Consulté le 22 February 2021].

Zelt, C.A., Barton, P.J. (1998). Three-dimensional seismic refraction tomography: A comparison of two methods applied to data from the Faeroe Basin. [En ligne*Journal of Geophysical Research: Solid Earth*], 103(B4), 7187–7210. [En ligne]. Disponible à l'adresse : http://agupubs.onlinelibrary.wiley.com/doi/abs/10.1029/97JB03536 [Consulté le 16 December 2020].

Zelt, C.A., Smith, R.B. (1992). Seismic traveltime inversion for 2-D crustal velocity structure. [En ligne*Geophysical Journal International*], 108(1), 16–34. [En ligne]. Disponible à l'adresse : https://academic.oup.com/gji/article/108/1/16/661014/Seismic-traveltime-inversion-for-2-D-crustal [Consulté le 7 February 2017].